\documentclass[preprint,5p]{elsarticle}

\usepackage{graphicx,amssymb}
\usepackage{epsf,psfig}

\journal{Chaos, Solitons \& Fractals}

\begin{document}

\begin{frontmatter}

\title{Are external perturbations responsible for chaotic motion in galaxies?}

\author{Euaggelos E. Zotos\corref{}}

\address{Department of Physics, \\
Section of Astrophysics, Astronomy and Mechanics, \\
Aristotle University of Thessaloniki \\
GR-541 24, Thessaloniki, Greece}

\cortext[]{Corresponding author: \\
\textit{E-mail address}: evzotos@astro.auth.gr (Euaggelos E. Zotos)}

\begin{abstract}

We study the nature of motion in a logarithmic galactic dynamical model, with an additional external perturbation. Two different cases are investigated. In the first case the external perturbation is fixed, while in the second case it is varying with the time. Numerical experiments suggest, that responsible for the chaotic phenomena is the external perturbation, combined with the dense nucleus. Linear relationships are found to exist, between the critical value of the angular momentum and the dynamical parameters of the galactic system that is, the strength of the external perturbation, the flattening parameter and the radius of the nucleus. Moreover, the extent of the chaotic regions in the phase plane, increases linearly as the strength of the external perturbation and the flattening parameter increases. On the contrary, we observe that the percentage covered by chaotic orbits in the phase plane, decreases linearly, as the scale length of the nucleus increases, becoming less dense. Theoretical arguments are used to support and explain the numerically obtained outcomes. A comparison of the present outcomes with earlier results is also presented.

\end{abstract}


\end{frontmatter}

\section{Introduction}

Most galaxies in the Universe, are gravitationally bound to a number of other galaxies. These form a fractal-like hierarchy of clustered structures, with the smallest such association being termed groups. A group of galaxies is the most common type of galactic cluster and these formations contain the majority of the galaxies. In other words, galaxies interact with each other, most common in pairs or as triple systems. The Milky Way-Magellanic Clouds system, is a very good example of galactic interaction [1,8,10,16,20,24,27,28]. Another interesting system of interacting galaxies, is the Andromeda $M31$ galaxy, with its two small companion galaxies $M32$ and $NGC$ $205$. This interesting triple system was investigated, in an earlier paper, using a self consistent computer simulation code, with interesting results [26]. Moreover, the formation of spiral structure in a galaxy, as a result of the gravitational perturbation caused by a permanent companion, was studied in a previous work [25].

In this article, we shall use the gravitational potential
\begin{equation}
V_0(r,z)=\frac{\upsilon_0^2}{2} \ln \left[r^2 + \alpha z^2 + c^2 \right],
\end{equation}
in order to describe the motion in our galactic dynamical system. This potential is important for galactic dynamics and represents an elliptical galaxy, with a nucleus of radius $c$, which displays a flat rotation curve at large radii [2]. Here $r, z$ are the usual cylindrical coordinates. The flattening parameter $1 \leq \alpha \leq 2$, defines the axial ratio of the equipotential ellipsoids. Logarithmic potentials, have been frequently used by many researchers, over the last decades, in order to model galactic motion [6,12,19,21,22]. The parameter $0.01 \leq c \leq 0.25$ is the radius of the nucleus, while $\upsilon_0$ is a parameter used for the consistency of the galactic units. In order to keep things simple, we consider that our galaxy is subject to an external perturbation $V_1 = -\lambda r$, caused by a companion galaxy, where $\lambda > 0$ is the strength of the external perturbation. Thus the total potential is
\begin{equation}
V(r,z)=V_0 + V_1=\frac{\upsilon_0^2}{2} \ln \left[r^2 + \alpha z^2 + c^2 \right] - \lambda r.
\end{equation}

We use a system of galactic units, where the unit of length is $1 kpc$, the unit of time is $0.97746 \times 10^8 yr$ and the unit of mass is $2.325 \times 10^7 M_{\odot}$. The velocity and the angular velocity units are $10 km/s$ and $10 km/s/kpc$ respectively, while $G$ is equal to unity. The energy unit (per unit mass) is $100 (km/s)^2$. In the above units, we use the value $\upsilon_0 = 15 km/s$, while $\alpha$, $\lambda$ and $c$ are treated as parameters. We consider only bounded motion, therefore we take $\lambda \leq 21$.

As the total potential $V(r,z)$ is axially symmetric and the $L_z$ component of the angular momentum is conserved, we use the effective potential
\begin{equation}
V_{eff}(r,z)=\frac{L_z^2}{2r^2} + V(r,z),
\end{equation}
in order to study the character of motion in the meridian $r-z$ plane. The equations of motion are
\begin{eqnarray}
\ddot{r} &=& -\frac{\partial V_{eff}(r,z)}{\partial r}, \nonumber \\
\ddot{z} &=& -\frac{\partial V_{eff}(r,z)}{\partial z},
\end{eqnarray}
where the dot indicates derivative with respect to time. The corresponding Hamiltonian is written as
\begin{equation}
H=\frac{1}{2}\left(p_r^2 + p_z^2\right) + V_{eff}(r,z) = E,
\end{equation}
where $p_r$ and $p_z$ are the momenta, per unit mass, conjugate to $r$ and $z$ respectively, while $E$ is the numerical value of the test particle's energy, which is conserved. Equation (5) is an integral of motion, which indicates that the total energy of the test particle (star), is conserved.

The outcomes of the present research are based on the numerical integration of the equation of motion (4). We use a Bulirsh-St\"{o}er integration routine in Fortran 95, with double precision in all subroutines. The accuracy of our results was checked by the constancy of the energy integral (5), which was conserved up to the twelfth significant figure.

Our aim is to connect the external perturbation caused by a companion galaxy, with the character of motion (regular or chaotic). Moreover, we try to find relationships, in order to connect the dynamical parameters of the system, that is the strength of the external perturbation, the flattening parameter and the radius of the nucleus, with the chaotic percentage and the critical value of the angular momentum $L_{zc}$. Before doing this, it would be interesting to try to relate, the chaotic orbits or the chaotic motion, in general, with actual astronomical issues.

One of the main reasons for the transition from ordered to chaotic stellar motion, is the presence of a massive object in the central regions of galaxies. Stars reaching the center on highly eccentric radial orbits, are scattered out of the galactic plane, displaying chaotic motion [23]. Furthermore, a central mass concentration can strongly perturb the stellar orbits in elliptical galaxies, which become chaotic [18]. Observational indications suggesting, the presence of strong central mass concentration, with a very sharply raising rotation curve. A second reason, for chaotic behavior in galactic motion, is the presence of strong external perturbations. As the perturbation increases, it destroys the stability of orbits and increases the amount of the stochasticity of the dynamical system. Resonances are also responsible for chaotic motion [9,11]. Here we must emphasize, that resonances are not only responsible for the presence of chaos in galaxies, but also for the chaotic motion in the solar system [14,29]. The reader can find interesting information on the chaotic motion in galaxies and its connection with observations, in the work of Grosb\o l [13].

The layout of this article is organized as follows. In Section 2, we study the structure of the phase plane and derive numerically relationships between the percentage of the area $A\%$ covered by the chaotic orbits in the phase plane and the parameters $\lambda$, $\alpha$ and $c$. In the same Section, we present relationships between the critical value of the angular momentum and the same basic parameters of the dynamical system. In Section 3, these relationships are also reproduced and explained, using theoretical arguments. In Section 4, we follow the evolution of the orbits, as the flattening parameter, or the strength of the external perturbation changes with time. We close with Section 5, where a discussion and the conclusions of the present research are given.

\section{Structure of the dynamical system}

Figure 1a-d shows the $r-p_r, z=0, p_z > 0$ Poincar\'{e} phase plane, for the Hamiltonian (5) obtained by numerical integration of the equations of motion (4). Figure 1a shows the $r-p_r$ phase plane when: $\lambda =0, \alpha =1, L_z=10$ and $E=467$. Here we have the case of a spherical unperturbed galaxy and therefore, as we expected, the entire phase plane is occupied by invariant curves produced by regular orbits of the 1:1 resonance. Figure 1b shows the $r-p_r$ phase plane when: $\lambda =21, \alpha =1, L_z=10$ and $E=300$. In this case, the spherical galaxy is affected by the largest permissible value of the external perturbation $\lambda$. Here we observe, that the majority of orbits are regular orbits. The external perturbation has as a result a considerable chaotic layer, which is confined in the outer parts of the phase plane. Secondary resonances, if any, are negligible. Figure 1c shows the $r-p_r$ phase plane when: $\lambda =0, \alpha =1.9, L_z=10$ and $E=467$. Here we have the case of the unperturbed, non spherical galaxy. As one can see, again the majority of orbits are regular, while a very small chaotic layer in the outer parts of the phase plane is present. Figure 1d shows the $r-p_r$ phase plane when: $\lambda =21, \alpha =1.9, L_z=10$ and $E=300$. One observes a large chaotic sea, while there are also areas of regular motion. In addition to the above, one can also observe smaller islands of invariant curves, embedded in the chaotic sea, which are produced by secondary resonances. The values of all other parameters are: $\upsilon_0=15$ and $c=0.25$. The values of the energy $E$, were chosen so that in all phase planes $r_{max} \simeq 8$.
\begin{figure*}[!tH]
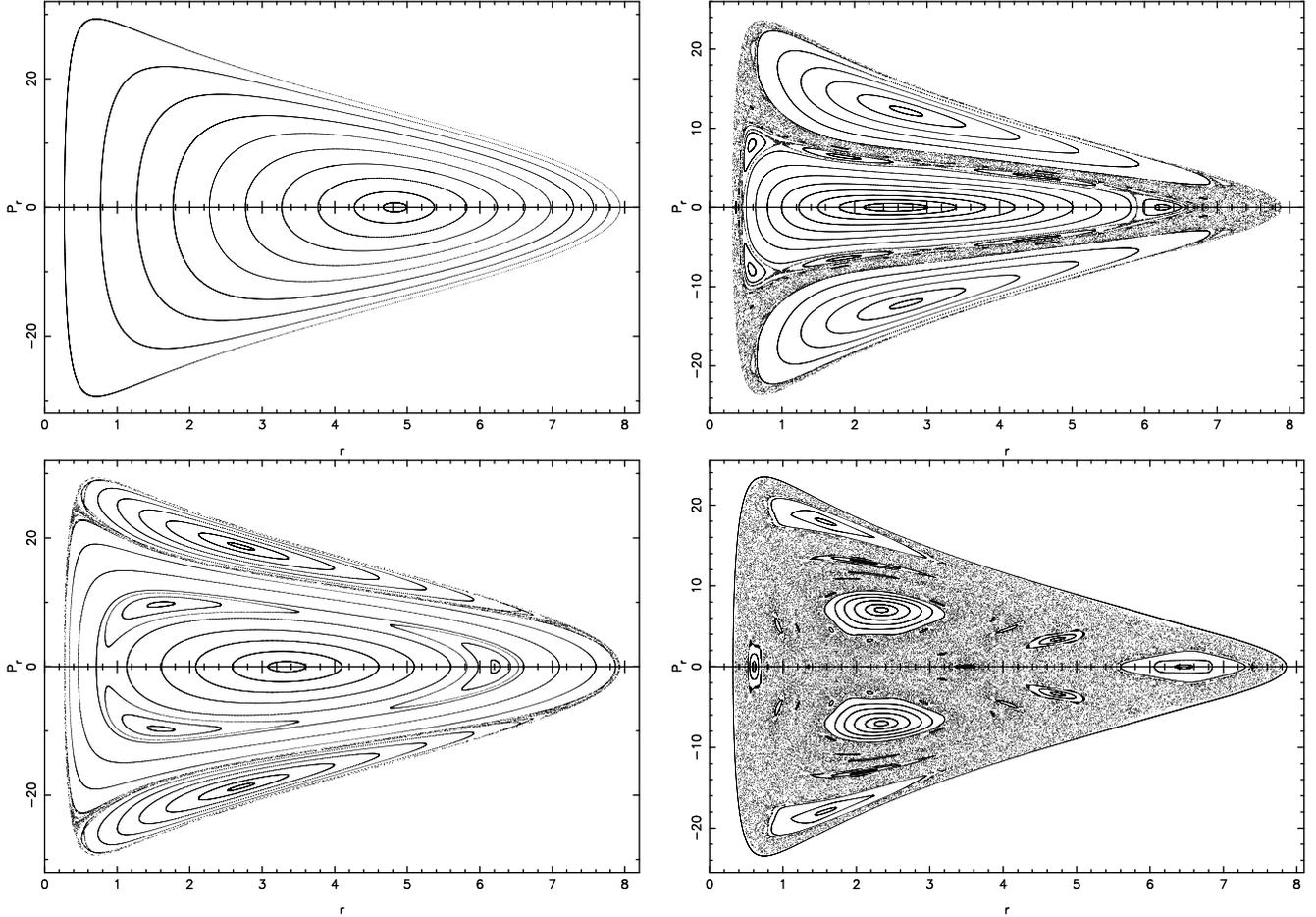

\centering
\resizebox{0.95\hsize}{!}{\rotatebox{270}{\includegraphics*{Fig-1a.ps}}\hspace{1cm}
                          \rotatebox{270}{\includegraphics*{Fig-1b.ps}}}
\resizebox{0.95\hsize}{!}{\rotatebox{270}{\includegraphics*{Fig-1c.ps}}\hspace{1cm}
                          \rotatebox{270}{\includegraphics*{Fig-1d.ps}}}
\vskip 0.1cm
\caption{(a-d): The $r-p_r,z=0, p_z>0$ Poincar\'{e} phase planes, for the Hamiltonian (5) when (a-upper left): $\lambda = 0, \alpha = 1, E=467$, (b-upper right): $\lambda = 21, \alpha = 1, E=300$, (c-down left): $\lambda = 0, \alpha = 1.9, E=467$ and (d-down right): $\lambda = 21, \alpha = 1.9, E=300$. The values of the other parameters are: $\upsilon_0 = 15, c=0.25$ and $L_z=10$.}
\end{figure*}
\begin{figure*}[!tH]
\resizebox{\hsize}{!}{\rotatebox{0}{\includegraphics*{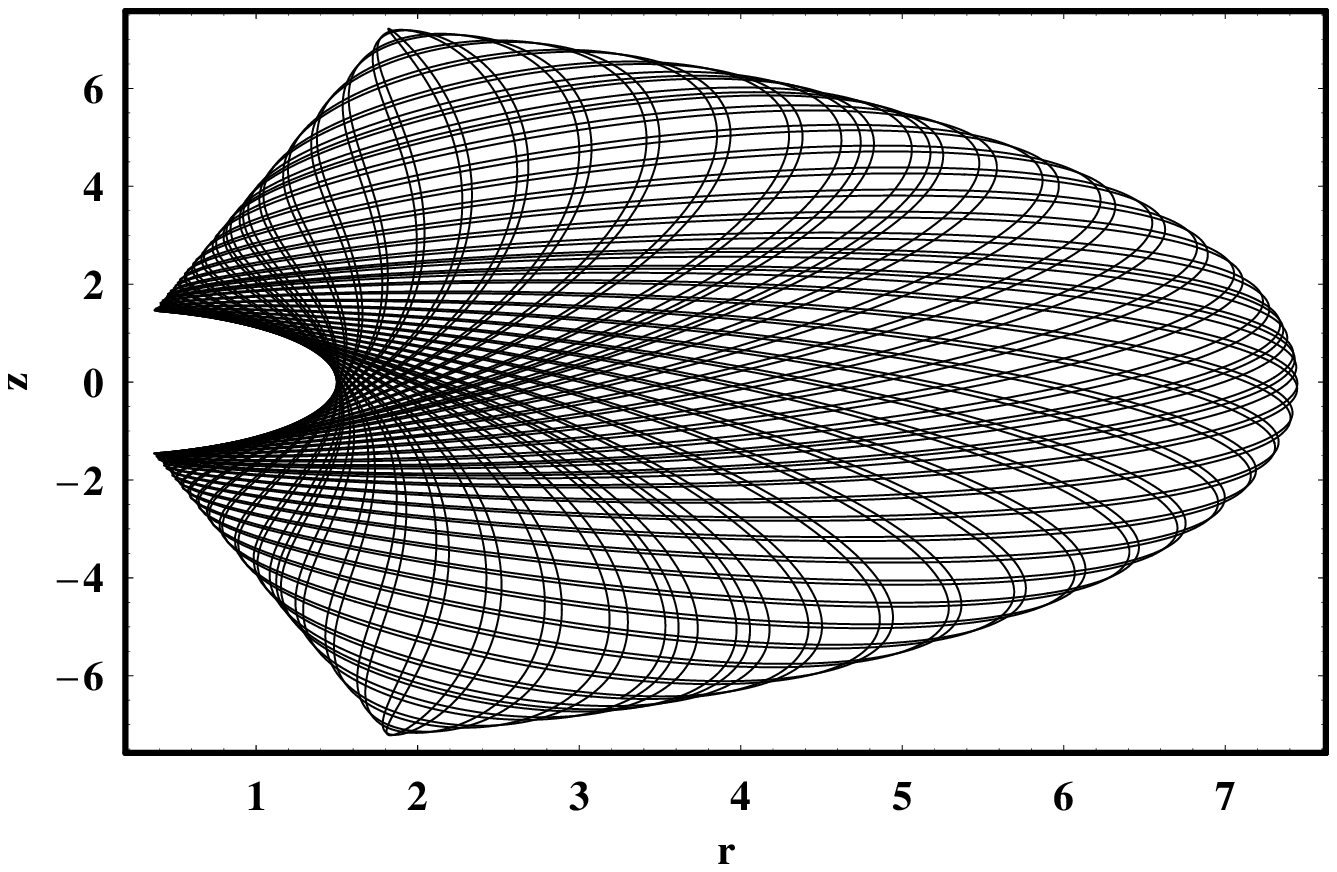}}\hspace{1cm}
                      \rotatebox{0}{\includegraphics*{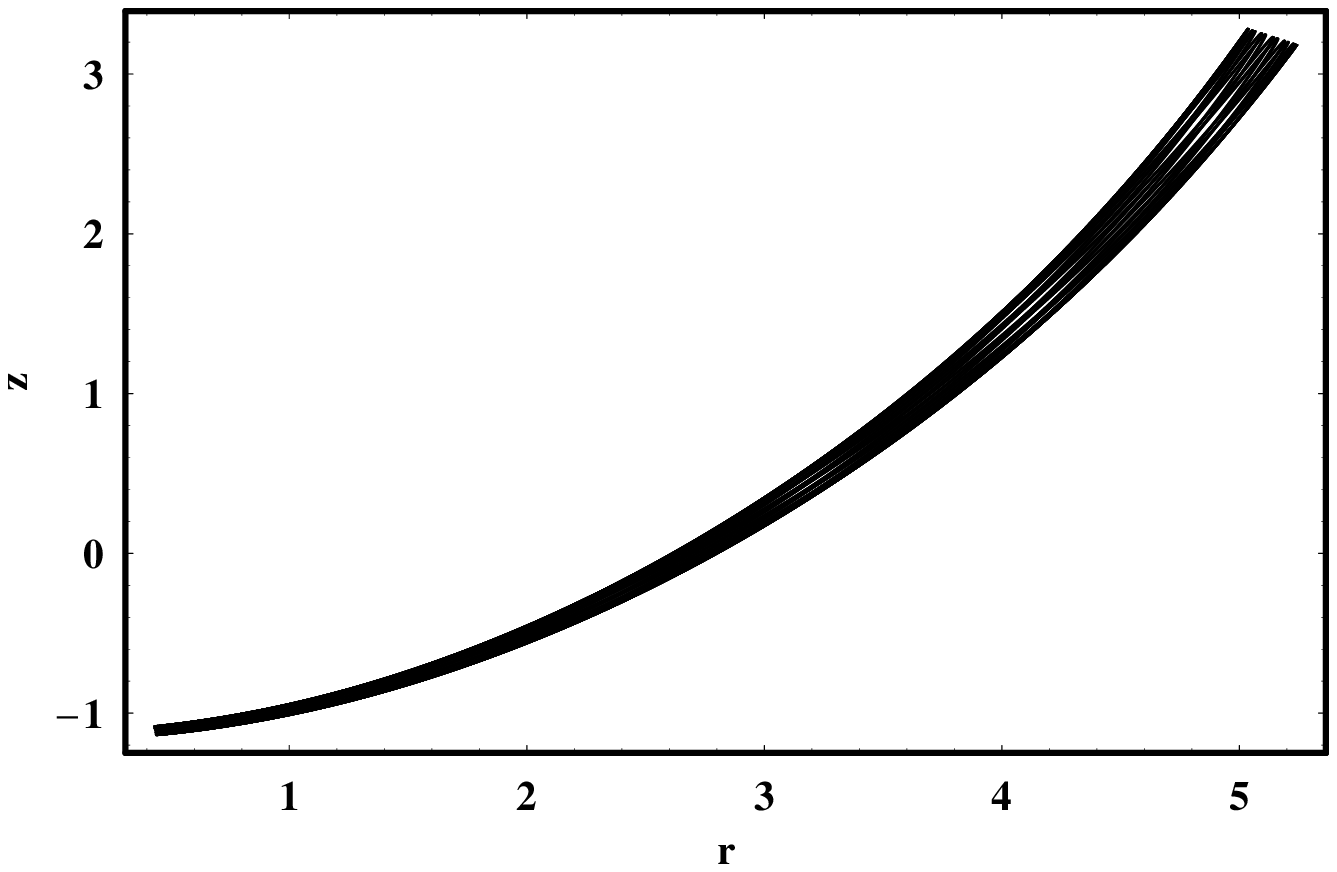}}}
\resizebox{\hsize}{!}{\rotatebox{0}{\includegraphics*{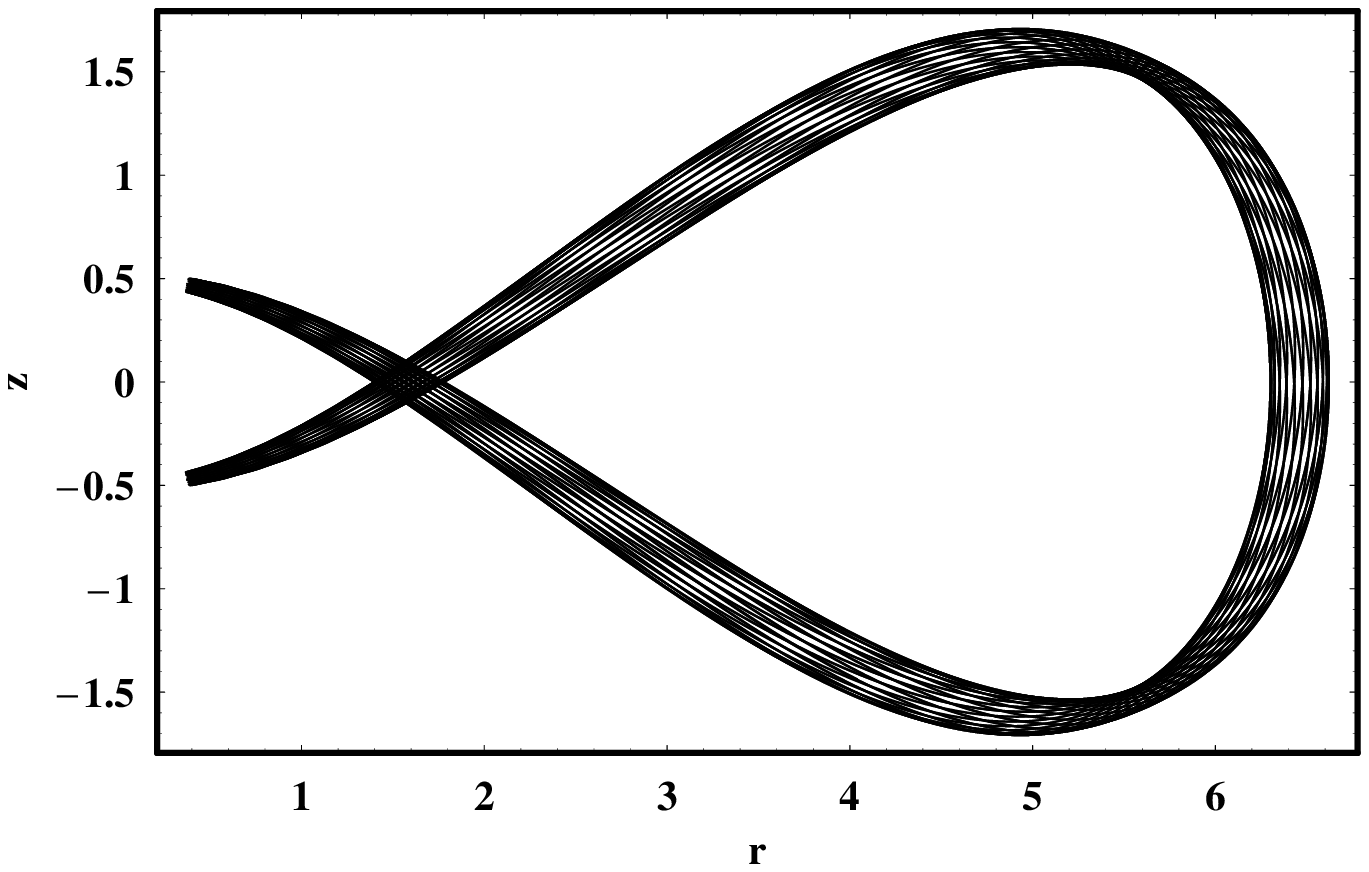}}\hspace{1cm}
                      \rotatebox{0}{\includegraphics*{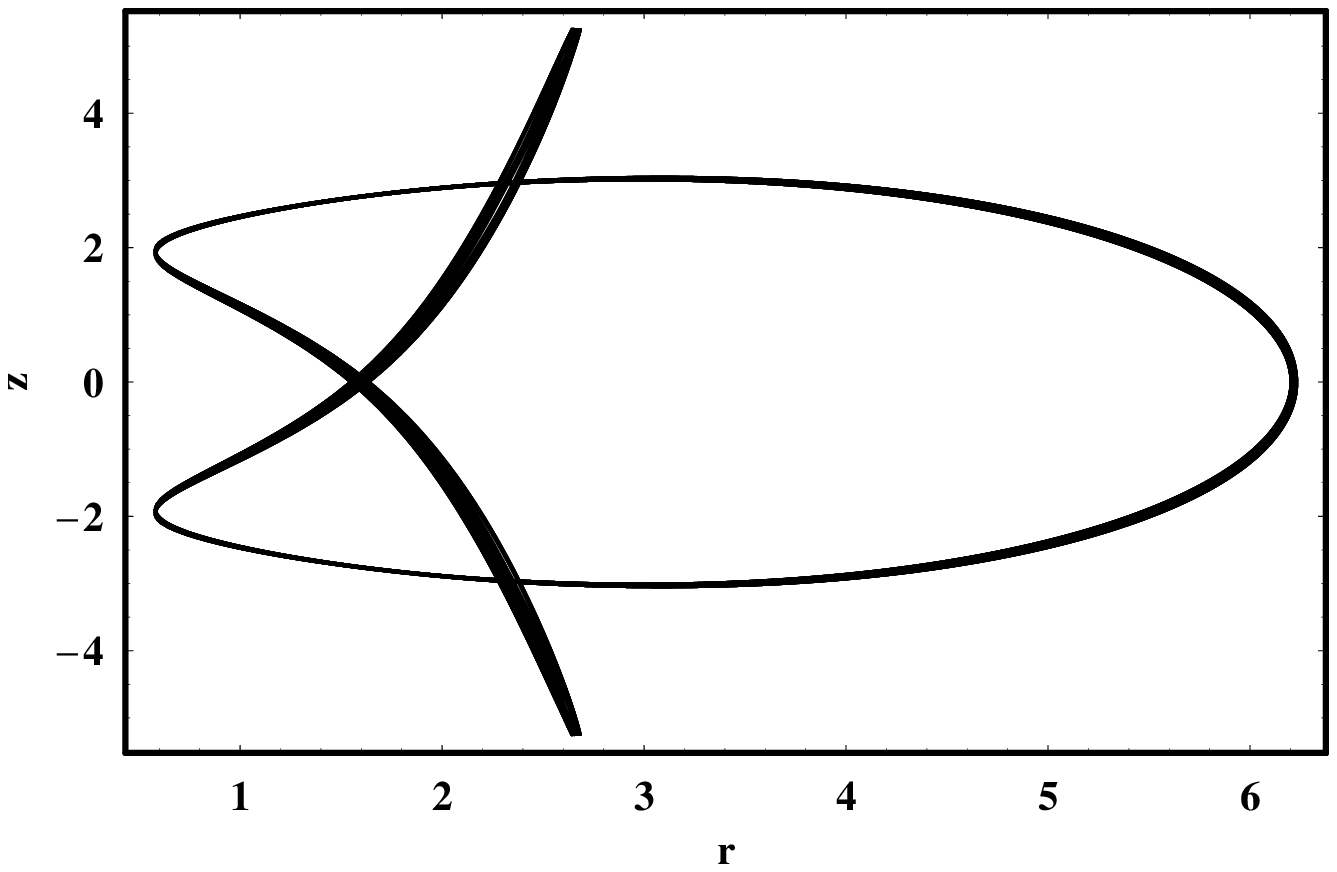}}}
\resizebox{\hsize}{!}{\rotatebox{0}{\includegraphics*{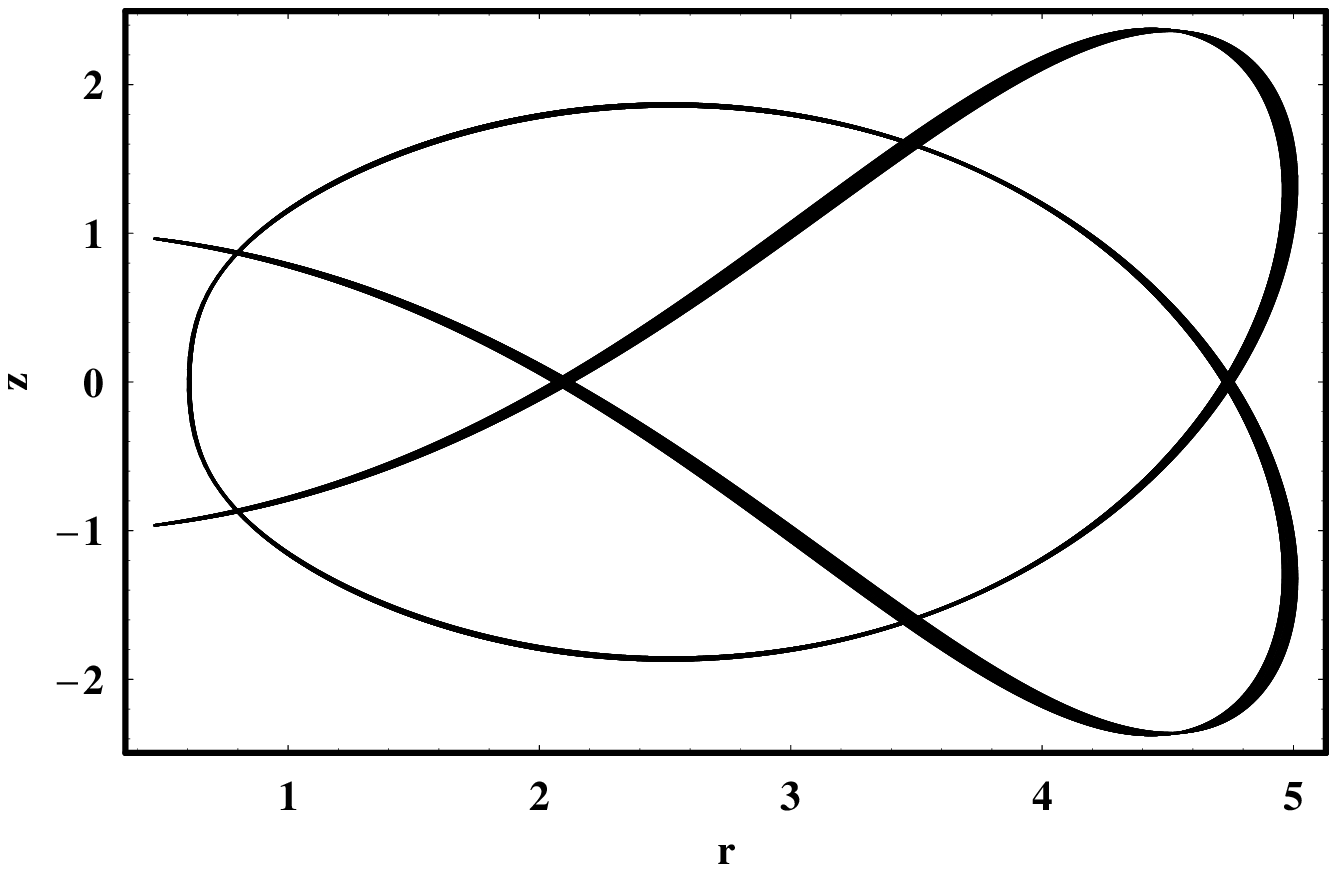}}\hspace{1cm}
                      \rotatebox{0}{\includegraphics*{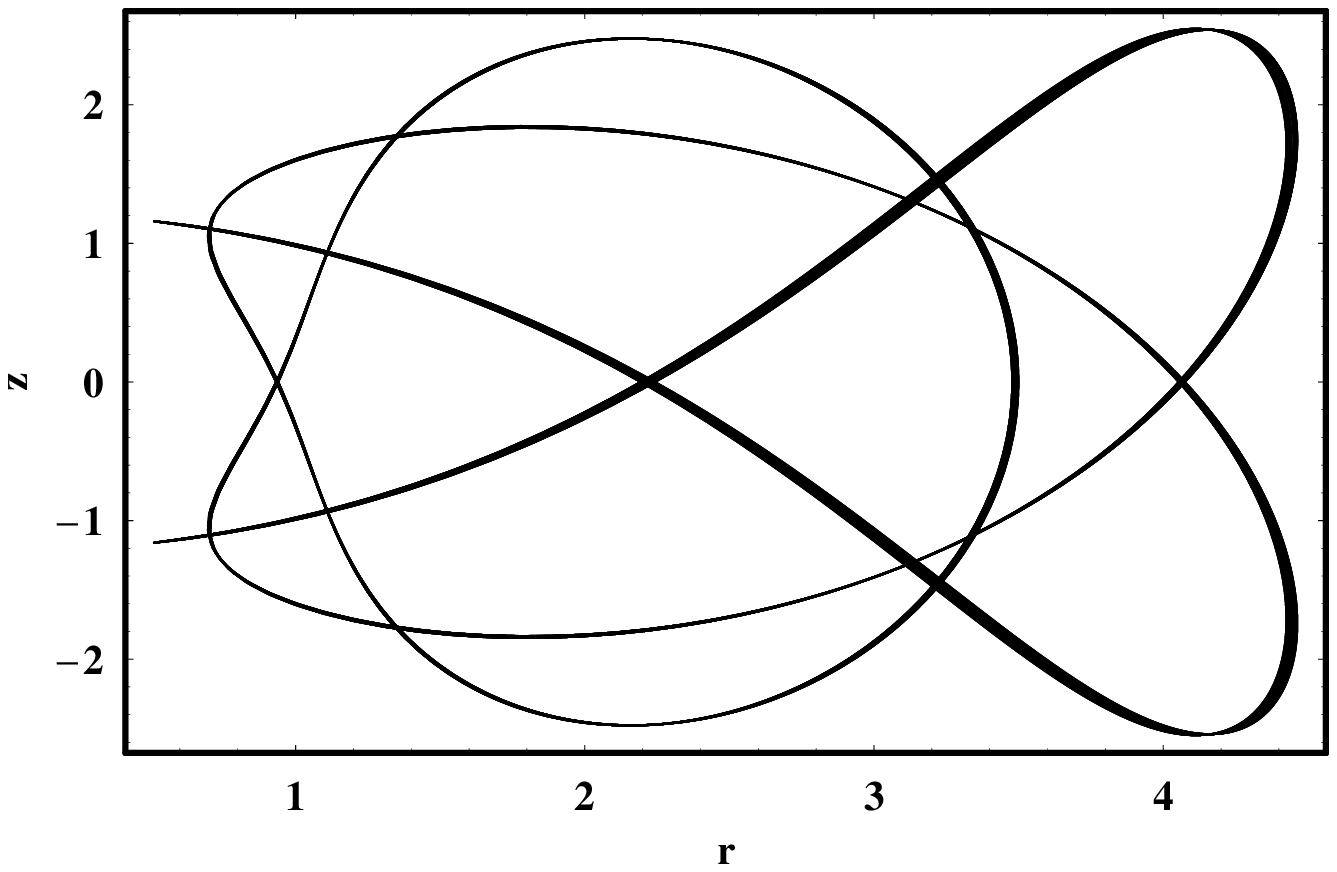}}}
\resizebox{\hsize}{!}{\rotatebox{0}{\includegraphics*{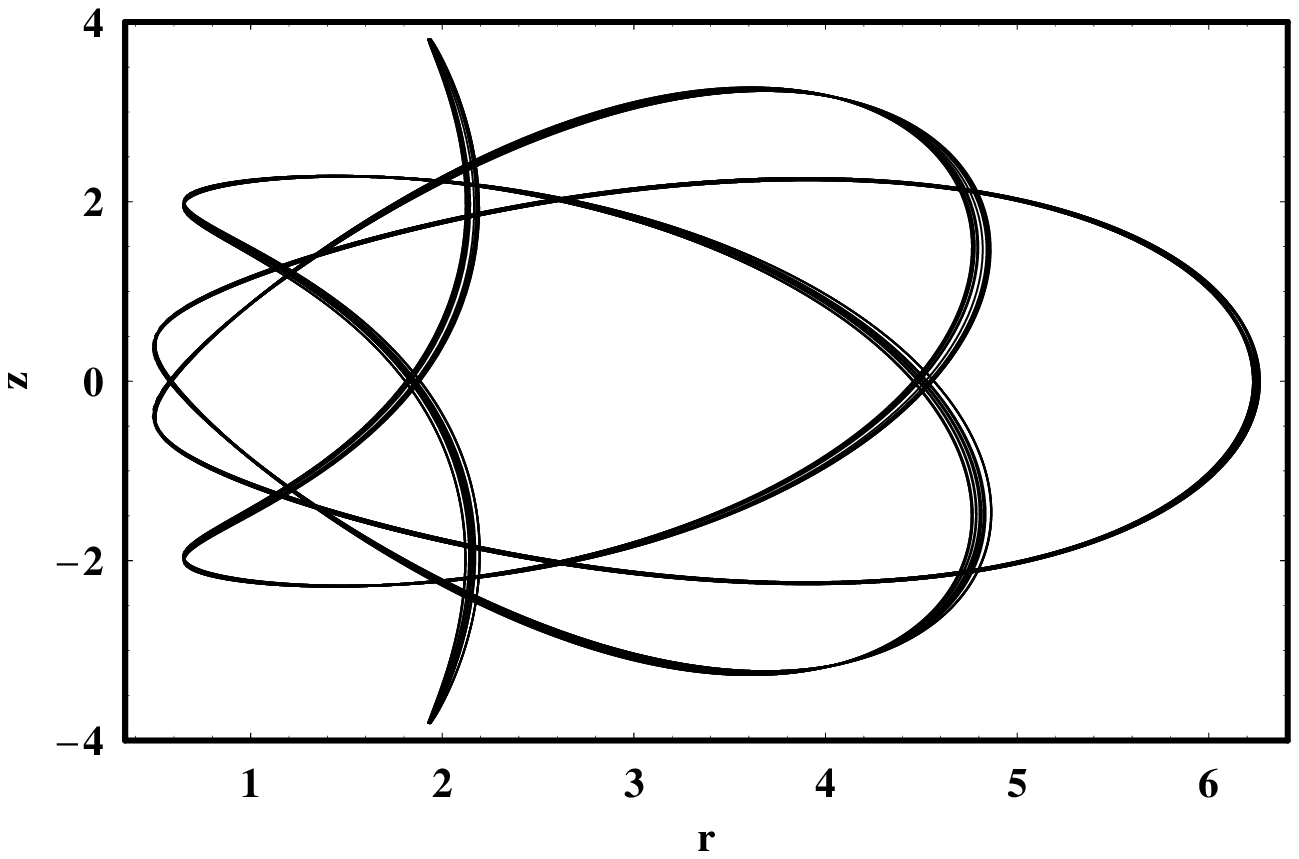}}\hspace{1cm}
                      \rotatebox{0}{\includegraphics*{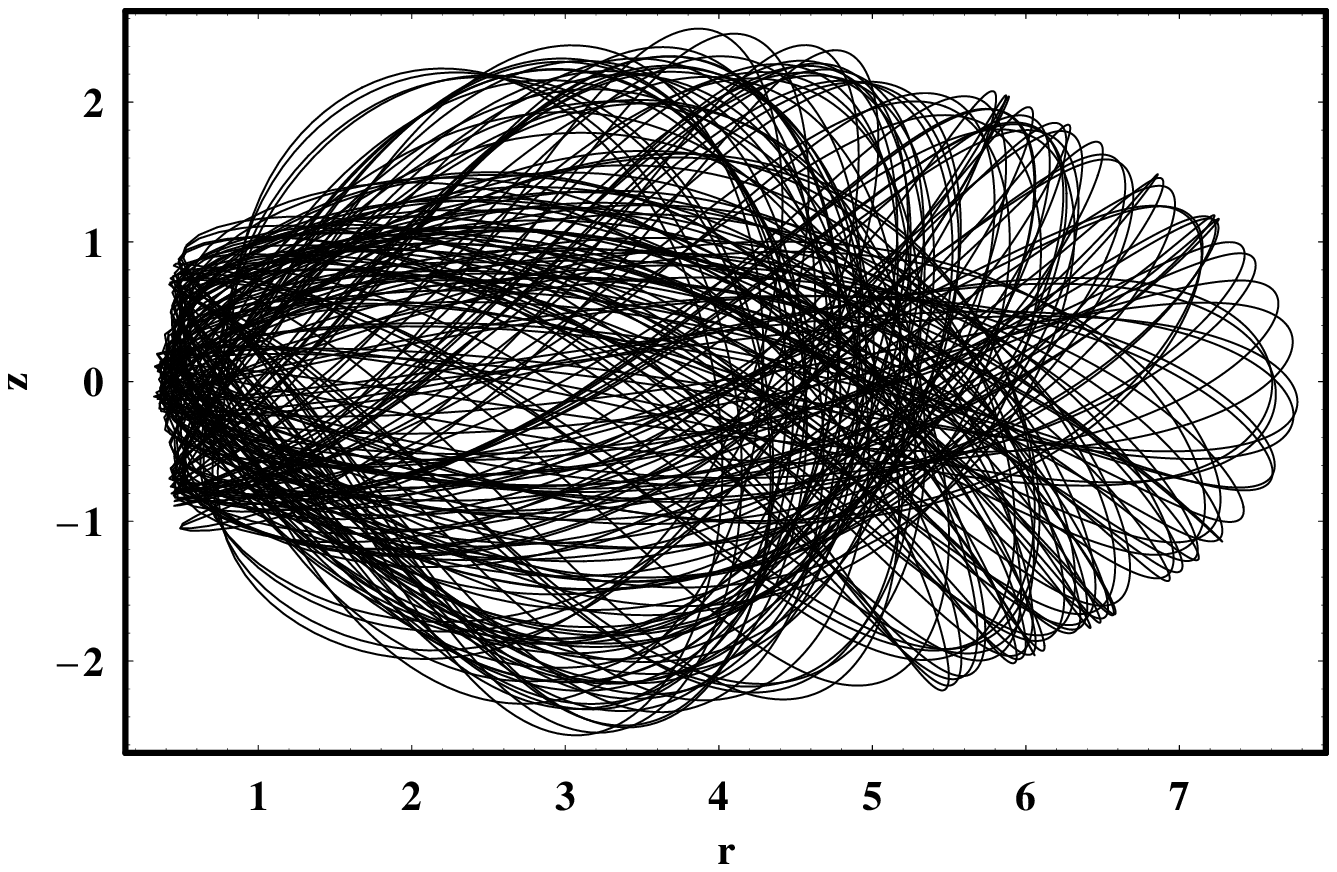}}}
\vskip 0.1cm
\caption{(a-h): Representative orbits of the dynamical system. Details are given in the text.}
\end{figure*}

The main conclusion from the above analysis, is that spherical galaxies display less chaos, than flat ones, when exposed to external perturbations. In other words, we see that chaos decreases when the symmetry of the galaxy increases.

Figure 2a-h shows eight representative orbits of the dynamical system. Figure 2a shows a regular orbit when: $\lambda =0, \alpha =1$ and $E=467$. Initial conditions are: $r_0=1.5, z_0=0, p_{r0}=0$, while the value of $p_{z0}$ is found from the energy integral (5), for all orbits. Figure 2b shows a regular orbit when: $\lambda =21, \alpha =1$ and $E=300$. Initial conditions are: $r_0=2.8, z_0=0, p_{r0}=12$. In Figure 2c a quasi periodic orbit, characteristic of the 2:3 resonance, is shown. Here: $\lambda =21, \alpha =1.9$ and $E=300$. Initial conditions are: $r_0=6.3, z_0=0, p_{r0}=0$. Figure 2d shows a quasi periodic orbit when: $\lambda =0, \alpha =1.9$ and $E=467$. Initial conditions are: $r_0=6.2, z_0=0, p_{r0}=0$. This orbit is characteristic of the 4:3 resonance. In Figure 2e a quasi periodic orbit, characteristic of the 4:5 resonance, is presented. Here: $\lambda =21, \alpha =1.9$ and $E=300$ while, initial conditions are: $r_0=0.6, z_0=0, p_{r0}=0$. In Figure 2f we see a characteristic orbit of the 6:7 resonance. Here: $\lambda =21, \alpha =1.9$ and $E=300$ while, initial conditions are: $r_0=3.5, z_0=0, p_{r0}=0$. Figure 2g shows a complicated orbit produced by the 8:7 resonance, when: $\lambda =21, \alpha =1$ and $E=300$. Initial conditions are: $r_0=6.23, z_0=0, p_{r0}=0$. A chaotic orbit is given in Figure 2h, when: $\lambda =21, \alpha =1.9$ and $E=300$. Initial conditions are: $r_0=0.42, z_0=0, p_{r0}=0$. All orbits were calculated for a time period of 100 time units. The values of all other parameters are: $\upsilon_0=15, c=0.25$ and $L_z=10$.
\begin{figure*}[!tH]
\centering
\resizebox{0.95\hsize}{!}{\rotatebox{0}{\includegraphics*{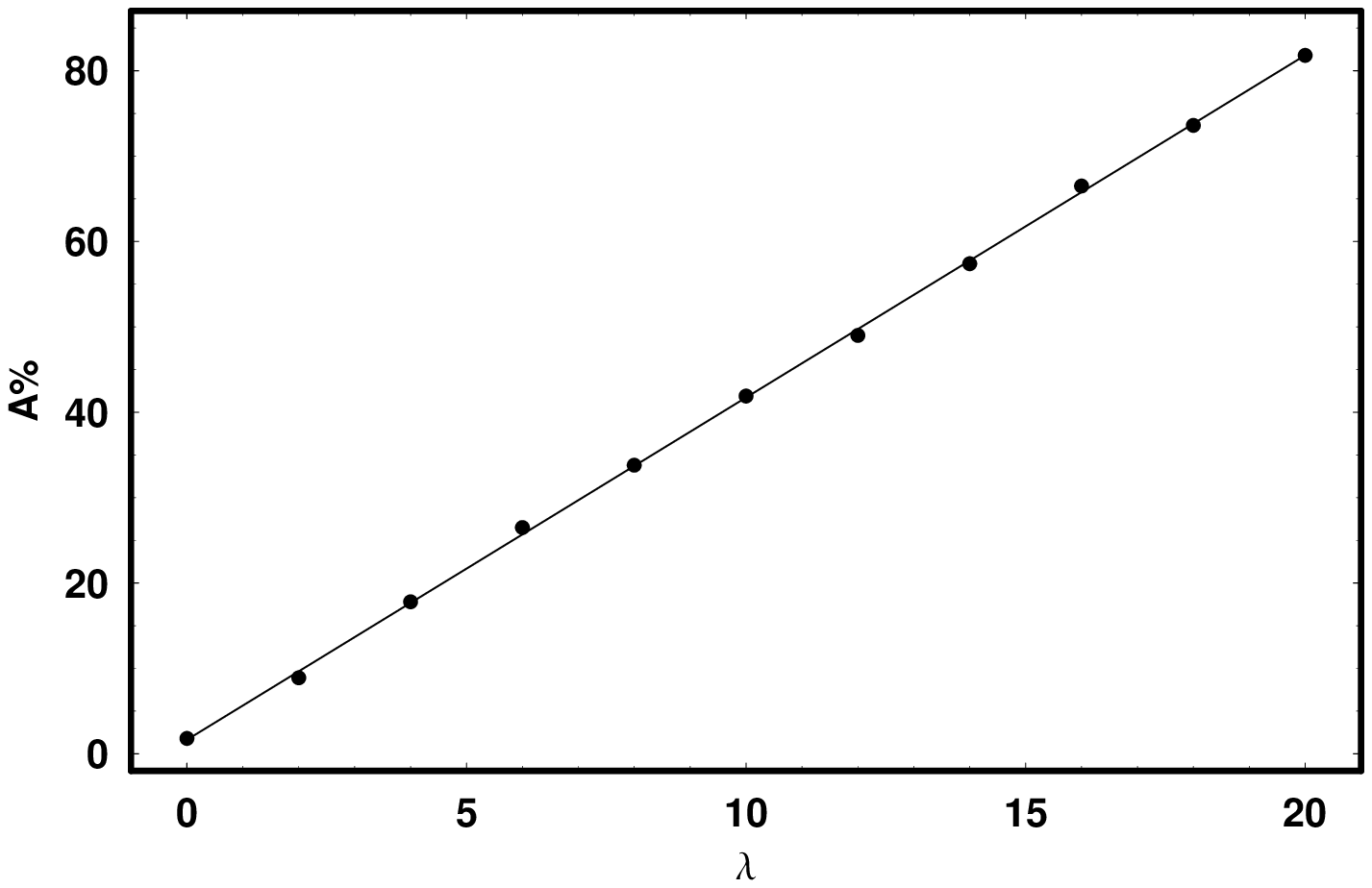}}\hspace{1cm}
                          \rotatebox{0}{\includegraphics*{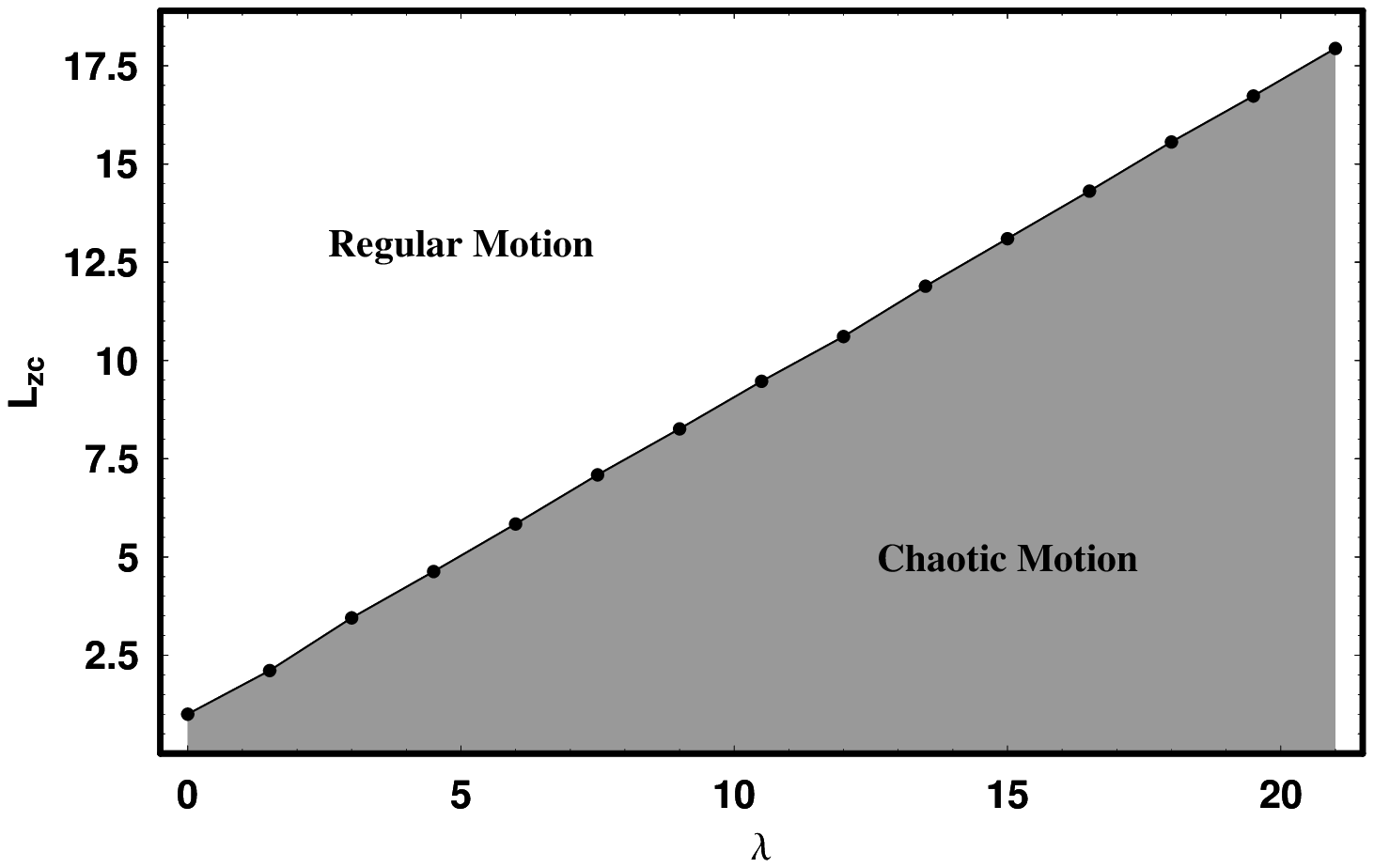}}}
\vskip 0.1cm
\caption{(a-b): (a-left): Plot of the percentage of the area $A\%$ in the phase plane covered by chaotic orbits vs. $\lambda$, when $L_z = 10$, (b-right): Relationship between the critical value of the angular momentum $L_{zc}$ and the external perturbation $\lambda$. The values of other parameters are: $\alpha = 1.9, c = 0.25$ and $E=300$.}
\end{figure*}
\begin{figure*}[!tH]
\centering
\resizebox{0.95\hsize}{!}{\rotatebox{0}{\includegraphics*{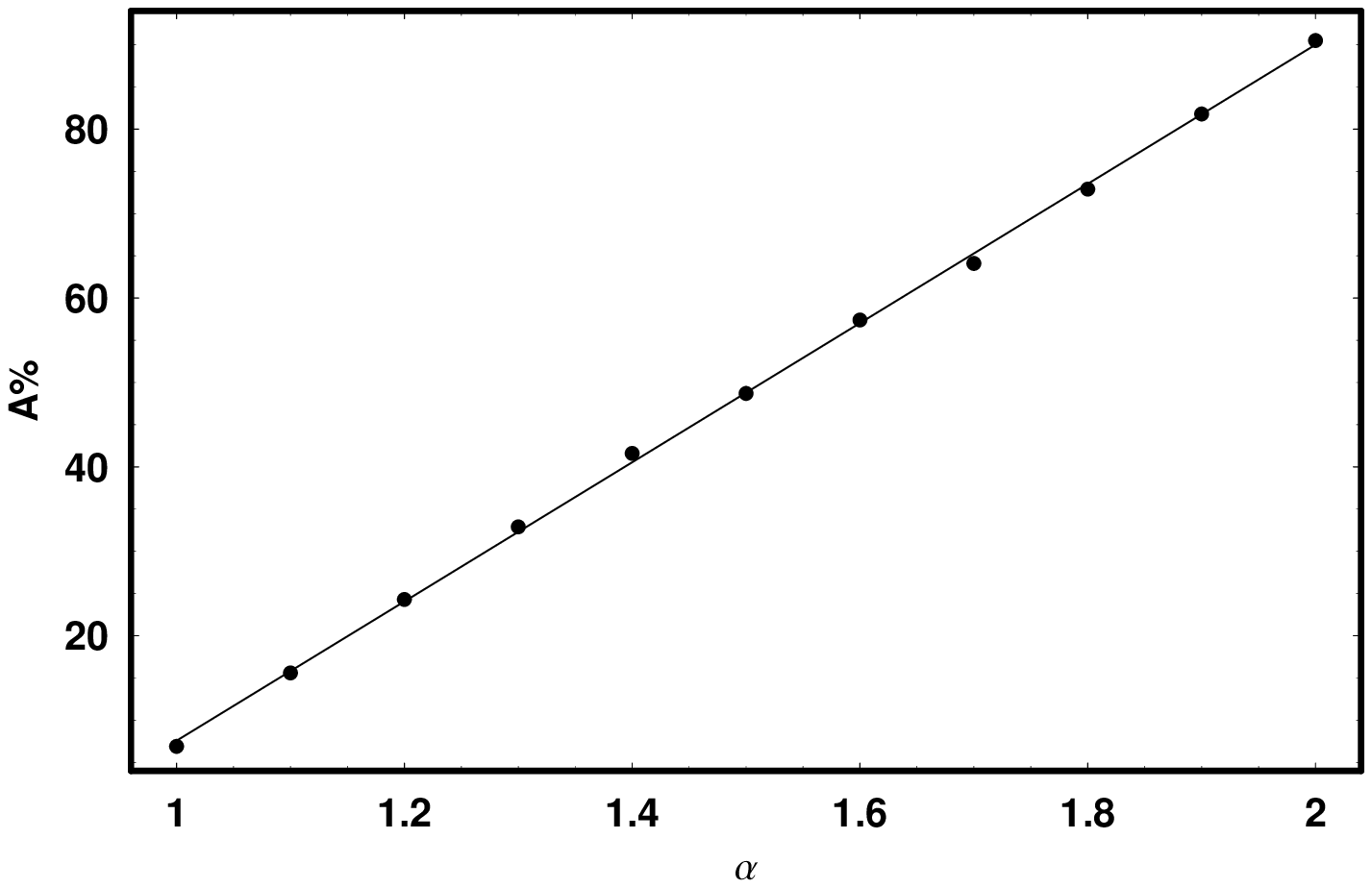}}\hspace{1cm}
                          \rotatebox{0}{\includegraphics*{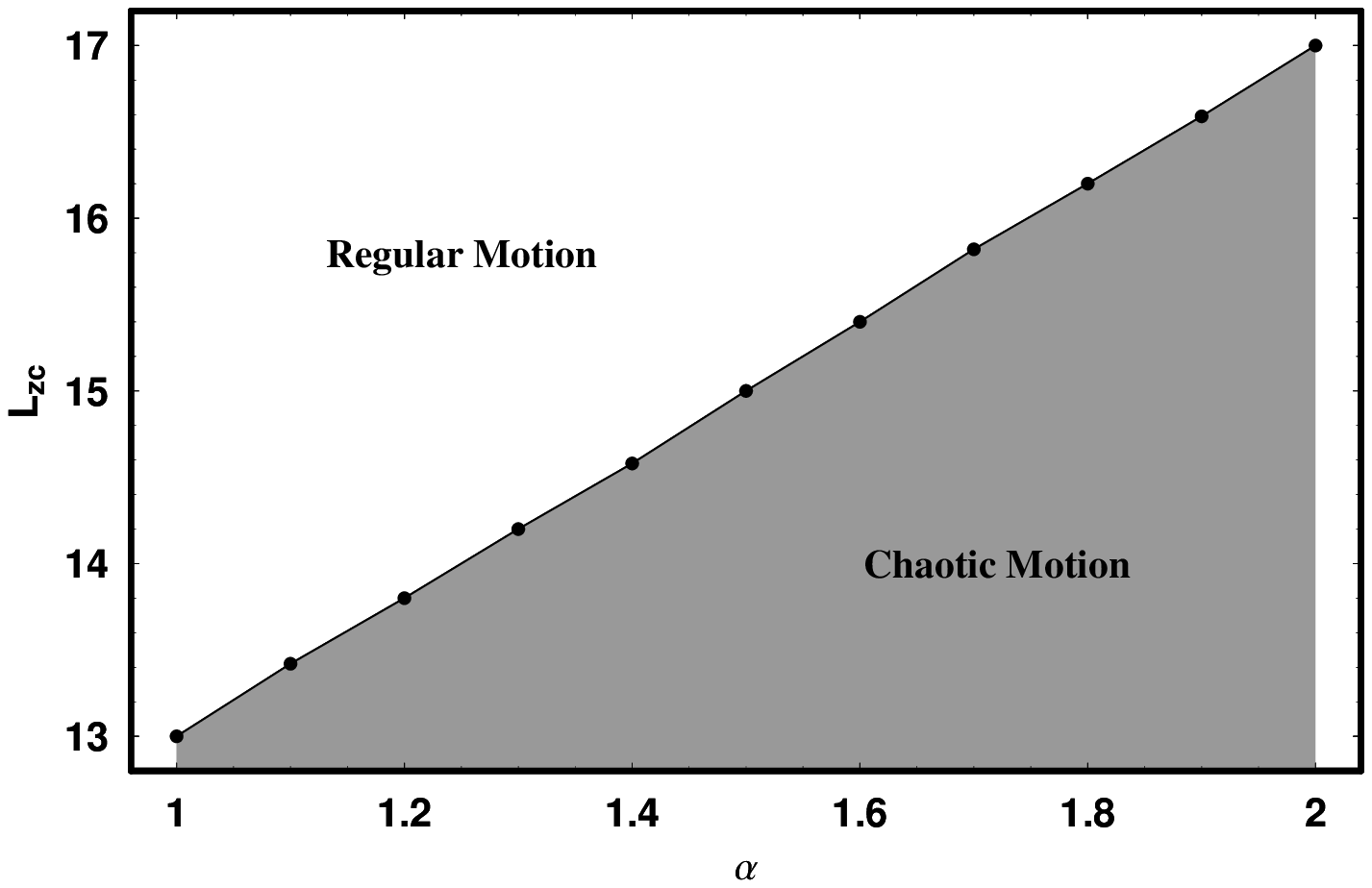}}}
\vskip 0.1cm
\caption{(a-b): (a-left): Plot of the percentage of the area $A\%$ in the phase plane covered by chaotic orbits vs. $\alpha$, when $L_z = 10$, (b-right): Relationship between the critical value of the angular momentum $L_{zc}$ and the flattening parameter $\alpha$. The values of other parameters are: $\lambda = 21, c = 0.25$ and $E=300$.}
\end{figure*}
\begin{figure*}[!tH]
\centering
\resizebox{0.95\hsize}{!}{\rotatebox{0}{\includegraphics*{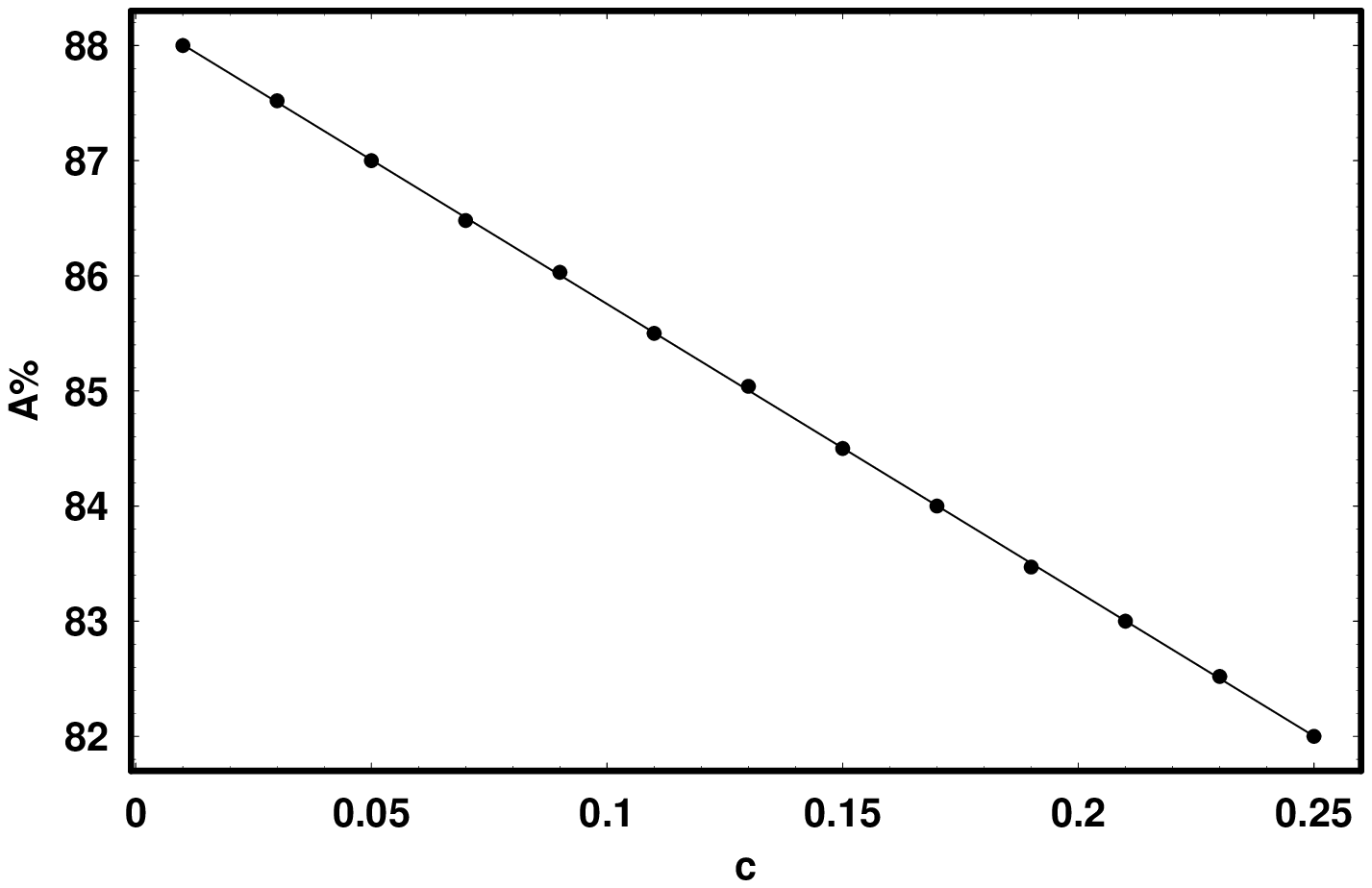}}\hspace{1cm}
                          \rotatebox{0}{\includegraphics*{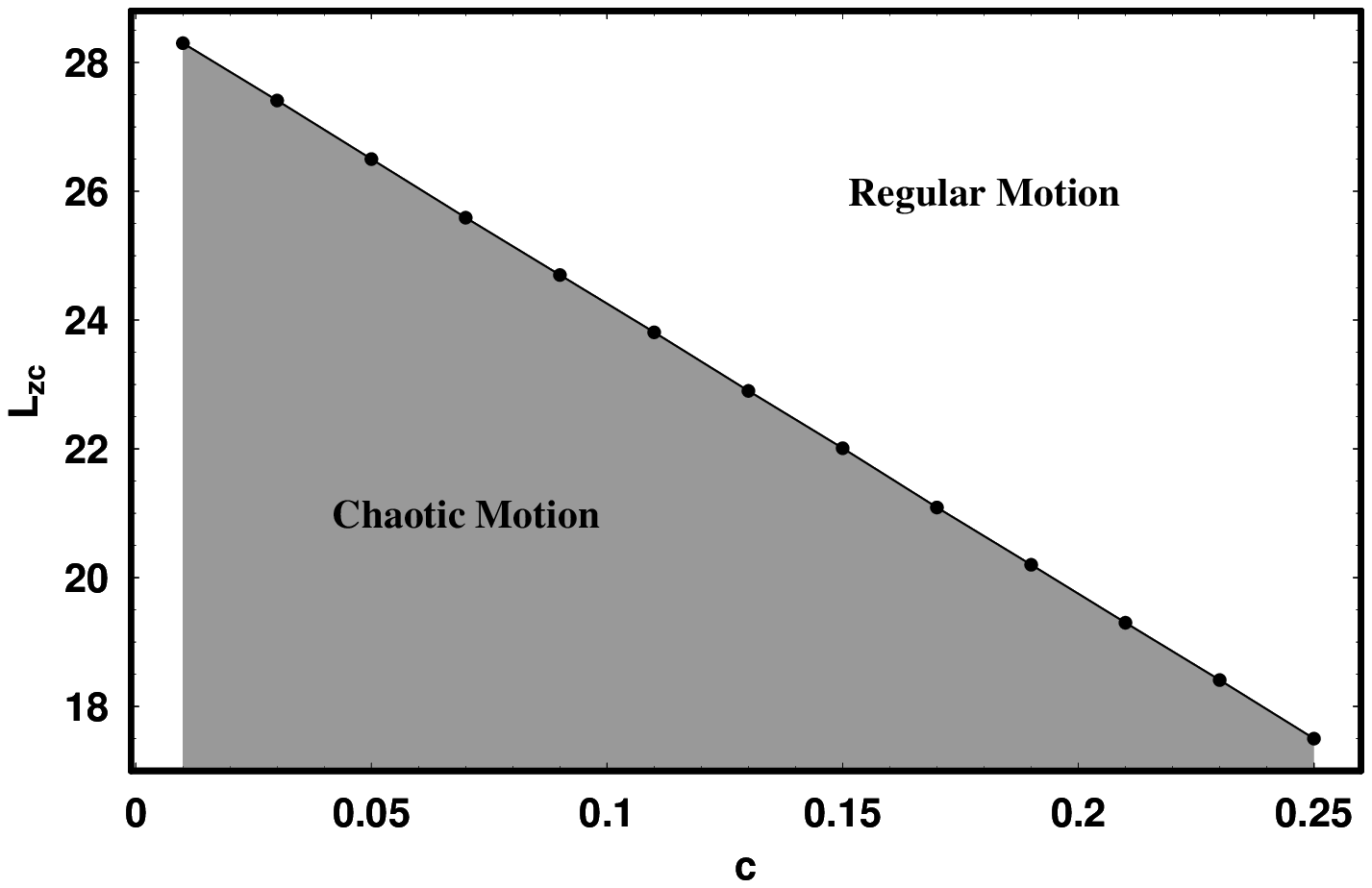}}}
\vskip 0.1cm
\caption{(a-b): (a-left): Plot of the percentage of the area $A\%$ in the phase plane covered by chaotic orbits vs. $c$, when $L_z = 10$, (b-right): Relationship between the critical value of the angular momentum $L_{zc}$ and the radius of the nucleus $c$. The values of other parameters are: $\alpha = 1.9, \lambda = 21$ and $E=300$.}
\end{figure*}

It is evident, that responsible for the chaotic regions is the external perturbation. This can be justified looking at Figure 3a, which displays the percentage of the area $A\%$ covered by chaotic orbits, in the $r-p_r, z=0, p_z>0$ phase plane as a function of the external perturbation $\lambda$. The values of all the other parameters are: $E = 300, \alpha = 1.9, c = 0.25$ and $L_z = 10$. We see that the chaotic area $A\%$ increases linearly with $\lambda$. Furthermore, the numerical calculations suggest that, for the above given values of the dynamical parameters, the chaotic regions are negligible when $\lambda \rightarrow 0$. Let us now come to see how the percentage of the chaotic regions in the $r-p_r$ phase plane, is connected with the flattening parameter $\alpha$. The results are given in Figure 4a, when $1 \leq \alpha \leq 2$. The values of all the other parameters are: $E = 300, \lambda = 21, c = 0.25$ and $L_z = 10$. One observes, that the chaotic area $A\%$ increases again linearly with $\alpha$. On the contrary, as we observe in Figure 5a, the percentage of the chaotic regions in the $r-p_r$ phase plane, decreases this time, linearly as the radius of the nucleus increases. The values of all other parameters are: $E = 300, \alpha = 1.9, \lambda = 21$ and $L_z = 10$. Therefore, we conclude that galaxies with less dense nucleus, display less chaotic motion. An explanation of these numerically found relationships of the dynamical system, will be given in the next Section.

Before closing this Section, we would like to present relationships connecting the critical value of the angular momentum $L_{zc}$ (that is the maximum value of the angular momentum $L_z$, for which the test particle displays chaotic motion) and the basic parameters of the dynamical system, that is the strength of the external perturbation $\lambda$, the flattening parameter $\alpha$ and the radius of the nucleus $c$. Our numerical experiments show that the relationship between $L_{zc}$ and $\lambda$ is linear. This linear relationship is shown in Figure 3b. In Figure 4b, we observe that the relationship connecting the flattening parameter $\alpha$ with the critical value of the angular momentum $L_{zc}$, ia also linear. Orbits starting in the upper part of the $\alpha - L_{zc}$ plane are regular, while orbits starting in the lower part of the same plane, including the line display chaotic motion. Figure 5b shows the relationship between the the critical value of the angular momentum $L_{zc}$ and the radius of the nucleus $c$. Here one can observe a straight line fitting all the numerically found data points. Orbits with values of $L_{zc}$ and $c$ on the left part of the $c - L_{zc}$ plane, including the line, are chaotic, while orbits with values of the parameters on the right part of the same plane are regular. What is more interesting, is that all these linear relationships can be also derived, using some elementary semi-theoretical arguments.

\section{Semi-theoretical arguments}

In this Section, we shall present some semi-theoretical arguments together with elementary numerical calculations, in order to explain the numerically obtained relationships given in Figs. 3, 4 and 5.

Our analysis takes place near the nucleus because, there, all internal forces acting on the star take their maximum values. Furthermore, the tangential velocity $\upsilon_{\phi}$ also takes its maximum value there. Note that the external force is constant, for all parts of the galaxy taking the value
\begin{equation}
F_{ext} = -\frac{\partial V_1}{\partial r} = \lambda.
\end{equation}

As the test particle approaches the nucleus there is a change in its momentum given by
\begin{equation}
m \Delta p_r = < F_{int} + F_{ext} > \Delta t = < F_{int} + \lambda > \Delta t,
\end{equation}
where $m$ is the mass of the test particle, $F_{int}$ and $F_{ext}$ is the internal and external average force acting on the particle, while $\Delta t$ is the duration of the encounter. We assume that the star displays chaotic motion, when the total change in the momentum after $n > 1$ encounters is of order of
\begin{equation}
m \upsilon_{\phi} = m \frac{L_{zc}}{<r_0>}, \ \ r_0 << 1.
\end{equation}

Thus we have
\begin{equation}
m \displaystyle\sum_{i=1}^{n} \Delta p_{ri} \approx < F_{int} + \lambda > \displaystyle\sum_{i=1}^{n} \Delta t_i.
\end{equation}

Setting
\begin{eqnarray}
m \displaystyle\sum_{i=1}^{n} \Delta p_{ri} &=& m \upsilon_{\phi} =
\frac{m L_{zc}}{<r_0>}, \nonumber \\
\displaystyle\sum_{i=1}^{n} \Delta t_i &=& T_c, \nonumber \\
m &=& 1,
\end{eqnarray}
in equation (9) we find
\begin{equation}
\frac{L_{zc}}{<r_0>} \approx < F_{int} + \lambda > T_c.
\end{equation}

Rearranging, we write equation (11) in the form
\begin{equation}
< L_{zc} > \approx L_{zc0} + \mu_1 \lambda,
\end{equation}
where  $\mu_1 = T_c r_0, L_{zc0} = F_{int} T_c r_0$. As the internal force, at a given point near the nucleus ($r = r_0 \ll c, z = z_0 \ll c$), is fixed, equation (12) gives the linear relationship between $\lambda$ and $L_{zc}$, which explains the numerical results shown in Figure 3b.

The linear relationship between the radius of the nucleus $c$ and $L_{zc}$ can explained similarly, using again semi-theoretical arguments. In fact, we use essentially similar arguments to those used in Caranicolas \& Innanen [3], Caranicolas \& Papadopoulos [5], Caranicoals \& Zotos [7] and Zotos [30]. When the star approaches the dense nucleus, its momentum in the $z$ direction changes according to the equation
\begin{equation}
m \Delta p_z = < F_z > \Delta t,
\end{equation}
where $m$ is the mass of the star, $< F_z >$ is the total average force acting in the $z$ direction, while $\Delta t$ is the duration of the encounter. It was assumed, that the star's deflection into higher values of $z$, proceeds in each time cumulatively, a little more, with each successive pass by the nucleus and not with a single tragic encounter. It is also assumed, that the star is scattered off the galactic plane, after $n (n>1)$ encounters, when the total change in the momentum in the $z$ direction, is of order of the tangential velocity $\upsilon_\phi$, of the star near the nucleus, at an average distance $r = < r_0 >$. Thus we have
\begin{equation}
m \displaystyle\sum_{i=1}^{n} \Delta p_{zi} \approx < F_z > \displaystyle\sum_{i=1}^{n} \Delta t_i.
\end{equation}

If we set
\begin{eqnarray}
m \displaystyle\sum_{i=1}^{n} \Delta p_{zi} &=& \frac{m L_{zc}}{< r_0 >}, \nonumber \\
\displaystyle\sum_{i=1}^{n} \Delta t_i &=& T_e, \nonumber \\
m &=& 1,
\end{eqnarray}
and combine equations (14) and (15), we find
\begin{equation}
\frac{L_{zc}}{< r_0 >} \approx < F_z > T_e.
\end{equation}
The force acting in the $z$ direction for a star of unit mass $(m=1)$ is
\begin{equation}
F_z = -\frac{\partial V_0}{\partial z} = \frac{- \alpha \upsilon_0^2 z}{r^2 + \alpha z^2 + c^2}.
\end{equation}

Setting $z=c, r = < r_0 > \gg z$ (remember that the star before scattering is very close to the galactic plane) and keeping only linear terms in $z$ and $c$, Eq. (17) becomes
\begin{equation}
< F_z > \approx - \frac{\alpha c}{< r_0 >^2}.
\end{equation}
Inserting the value from Eq. (18) into relation (16) we obtain
\begin{equation}
< L_{zc} > \approx - \frac{\alpha c}{< r_0 >} T_e = - \alpha \mu_2 c,
\end{equation}
where $\mu_2 = T_e/< r_0 >$. Here we must note that, equation (19) cannot be considered as an exact representation of the relation, between the involved quantities. It rather can be seen, as an indication of the relation that needs to be completed with additional terms. Those terms, can be derived through numerical calculations, providing the necessary information. Actually, numerical calculations, suggest that expression (19), needs to be supplemented with an additional constant term giving the value of $L_{zc}$ in the case when $c \rightarrow 0$. Calling this term $L_{zc0}$, expression (19) takes the form
\begin{equation}
< L_{zc} > \approx L_{zc0} - \alpha \mu_2 c.
\end{equation}

Relation (20) explains the linear relationship between the radius of the nucleus $c$ and $L_{zc}$, shown in Fig. 5b. Equation (20) also contains the flattening parameter $\alpha$. As one can see, for a given value of $c$, there is a linear dependence between $\alpha$ and $L_{zc}$. This explains the numerically obtained linear dependence shown in Fig. 4b, where $c = 0.25, \lambda = 21$ and $E = 300$.

As the scattering occurs near the nucleus, it must be $r_0 \ll c$ and $z_0 \ll c$. The particular values of $r_0$ and $z_0$ are irrelevant. As we can see from Eq. (17), $|F_z|$ increases linearly with the flattening parameter $\alpha$. As a consequence of this linear increase of the $|F_z|$, we see that the chaotic region in the $r-p_r$ phase plane, increase to high values, up to about $A = 82\%$, when $\alpha$ reaches 2 (see Fig. 4a).

\section{Evolution of orbits in the time-dependent model}

In this Section, we shall study the evolution of orbits as the parameters $\alpha$ and $\lambda$ change linearly with time following the equations
\begin{eqnarray}
\alpha(t) &=& \alpha_{in} - k_1 t, \nonumber \\
\lambda(t) &=& \lambda_{in} + k_2 t,
\end{eqnarray}
where $\alpha_{in}, \lambda_{in}$ are the initial values of $\alpha$ and $\lambda$, while $k_1$ and $k_2$ are
parameters.

Figure 6a-b shows the evolution of an orbit, as $\alpha$ changes with time, following the first of equations (21). The initial conditions are: $r_0 = 4, z_0 = p_{r0} = 0$. The initial value of energy is $E = 300$, $\lambda = 21, c = 0.25$, while the value of $p_{z0}$ is found from the energy integral (5) in all cases. The initial value of $\alpha$ is $1.9$, while $k_1$ is equal to $0.01$. The values of all other parameters are as in Fig. 1d. Figure 6a shows the orbit for the first $90$ time units, while figure 6b shows the orbit for the rest $110$ time units. One observes, that the orbit starts as chaotic and tends to be a regular orbit as $\alpha$ approaches unity. At $t = 90$ when $\alpha = 1$, the time evolution stops and the system is now spherical, but always subject to a large external perturbation $(\lambda = 21)$. The energy value was settled to the value $E = 276.48$. The evolution of the orbit in the spherically symmetric system is shown in Figure 6b. As expected, the orbit is now regular and remains regular. Of course, the value of energy $E$ must correspond to $r < 6$, for which regular orbits do exist. In Figure 6c one observes the complete evolution of the orbit for a time period of 200 time units. Figure 6d shows the evolution of the Lyapunov Characteristic Exponent (L.C.E) [17], for a time period of $10^5$ time units. For the first 90 time units, the profile of the L.C.E indicates chaotic motion, but for the rest time interval, the L.C.E corresponds to regular motion. Here we must point out, that the time scale with which the orbits become regular depends on the initial values of the quantities $\lambda$ and $c$. Numerical calculations not given here suggest that, as the system evolves from a flat system to a spherical one, the
majority of chaotic orbits become regular.
\begin{figure*}[!tH]
\centering
\resizebox{0.95\hsize}{!}{\rotatebox{0}{\includegraphics*{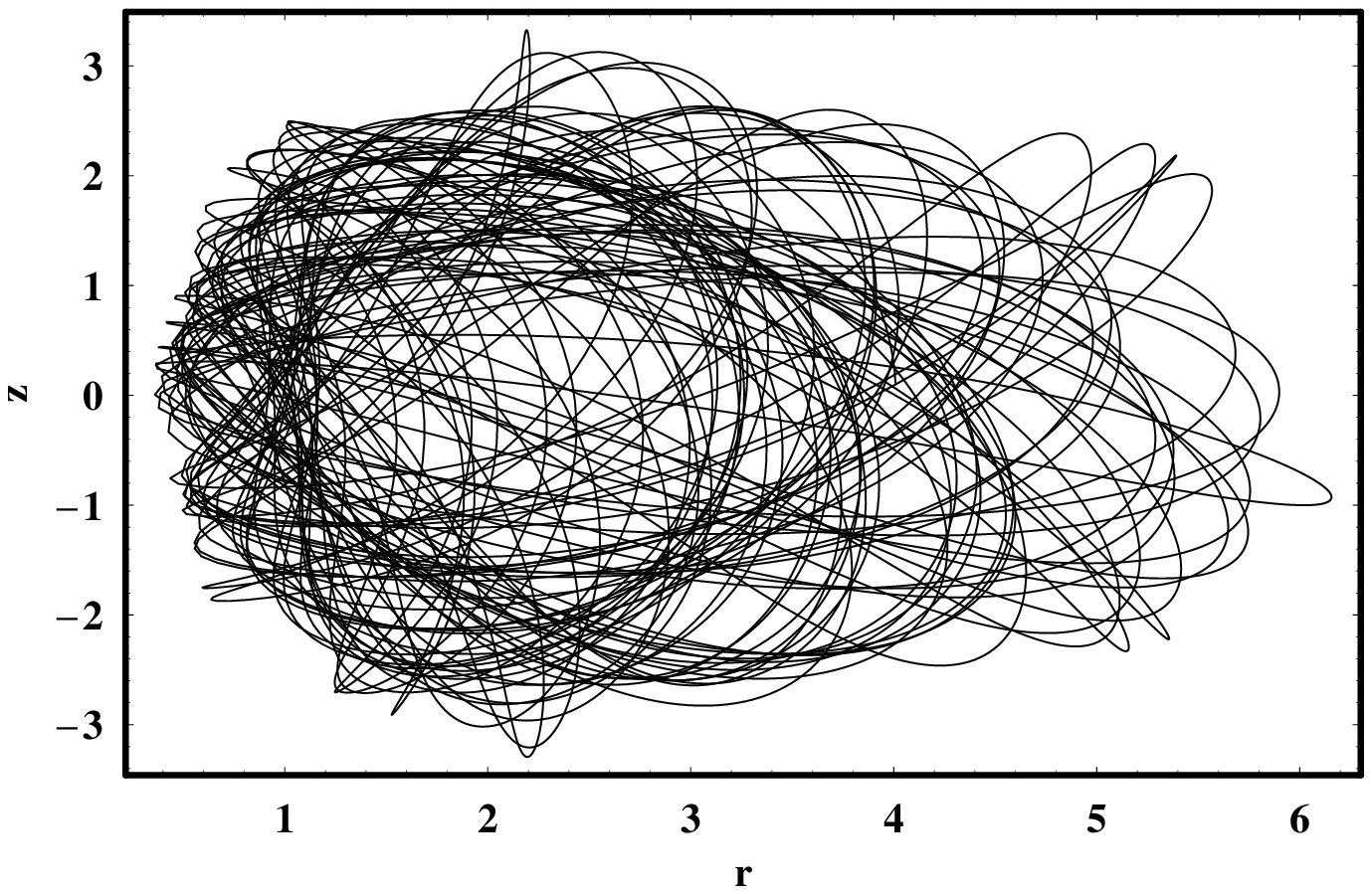}}\hspace{1cm}
                          \rotatebox{0}{\includegraphics*{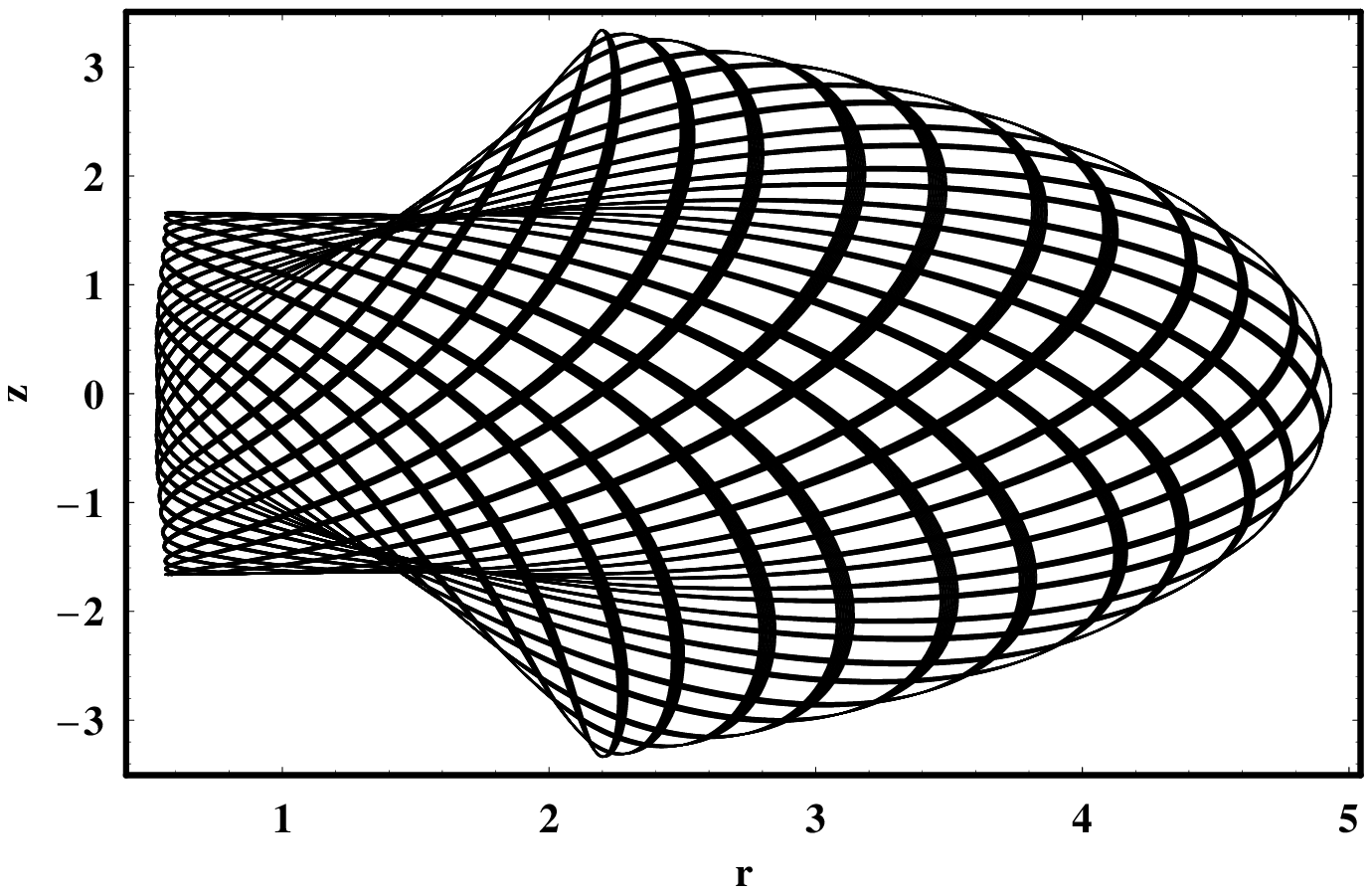}}}
\resizebox{0.95\hsize}{!}{\rotatebox{0}{\includegraphics*{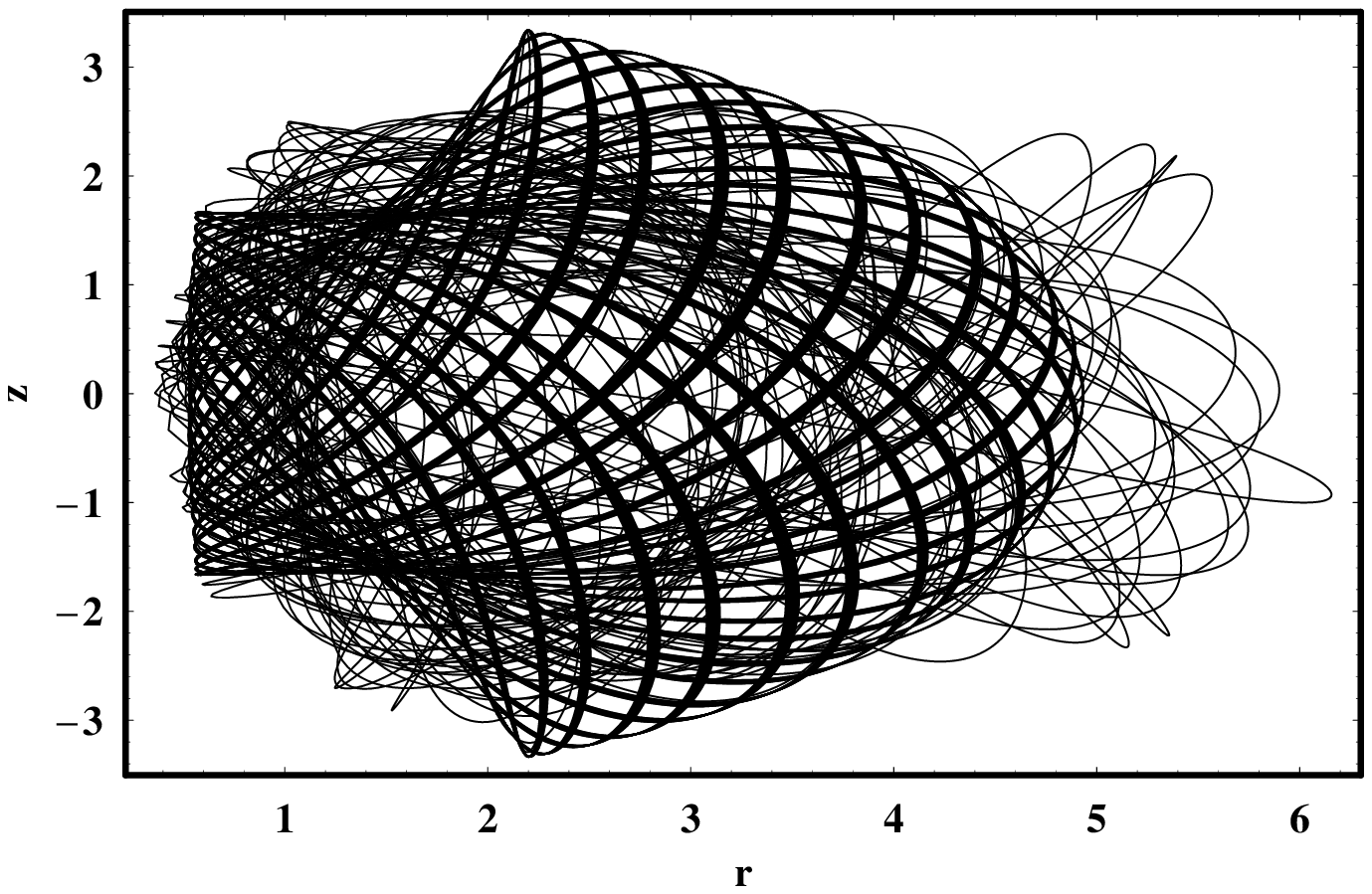}}\hspace{1cm}
                          \rotatebox{0}{\includegraphics*{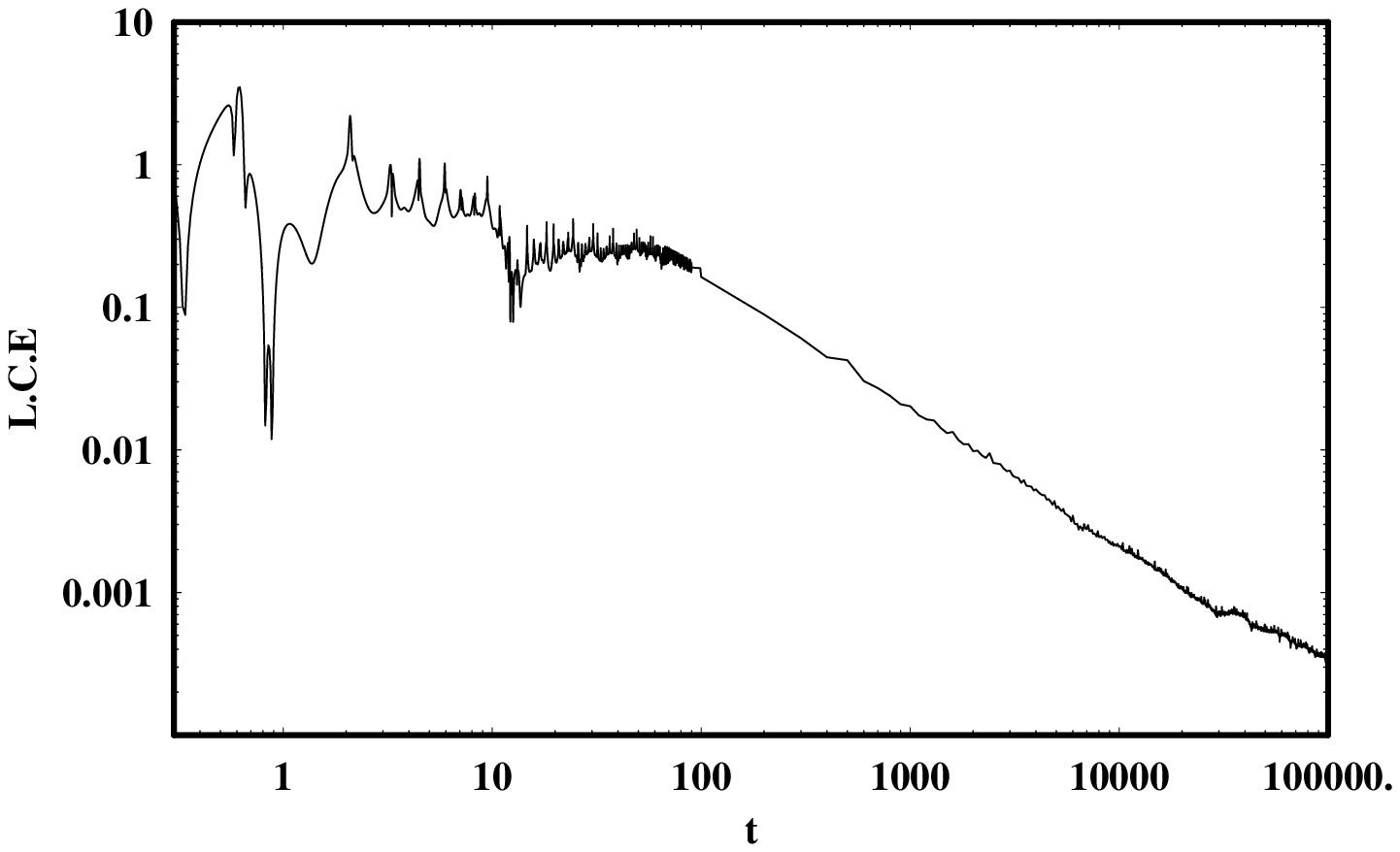}}}
\vskip 0.1cm
\caption{(a-d): (a-c): Evolution of an orbit as the flattening parameter $\alpha$ changes with time, following the first of equations (21). The orbit starts as chaotic and ends as a regular box orbit, in the spherical galaxy. (d-down right): Evolution of the L.C.E of this orbit. Details are given in the text.}
\end{figure*}
\begin{figure*}[!tH]
\centering
\resizebox{0.95\hsize}{!}{\rotatebox{0}{\includegraphics*{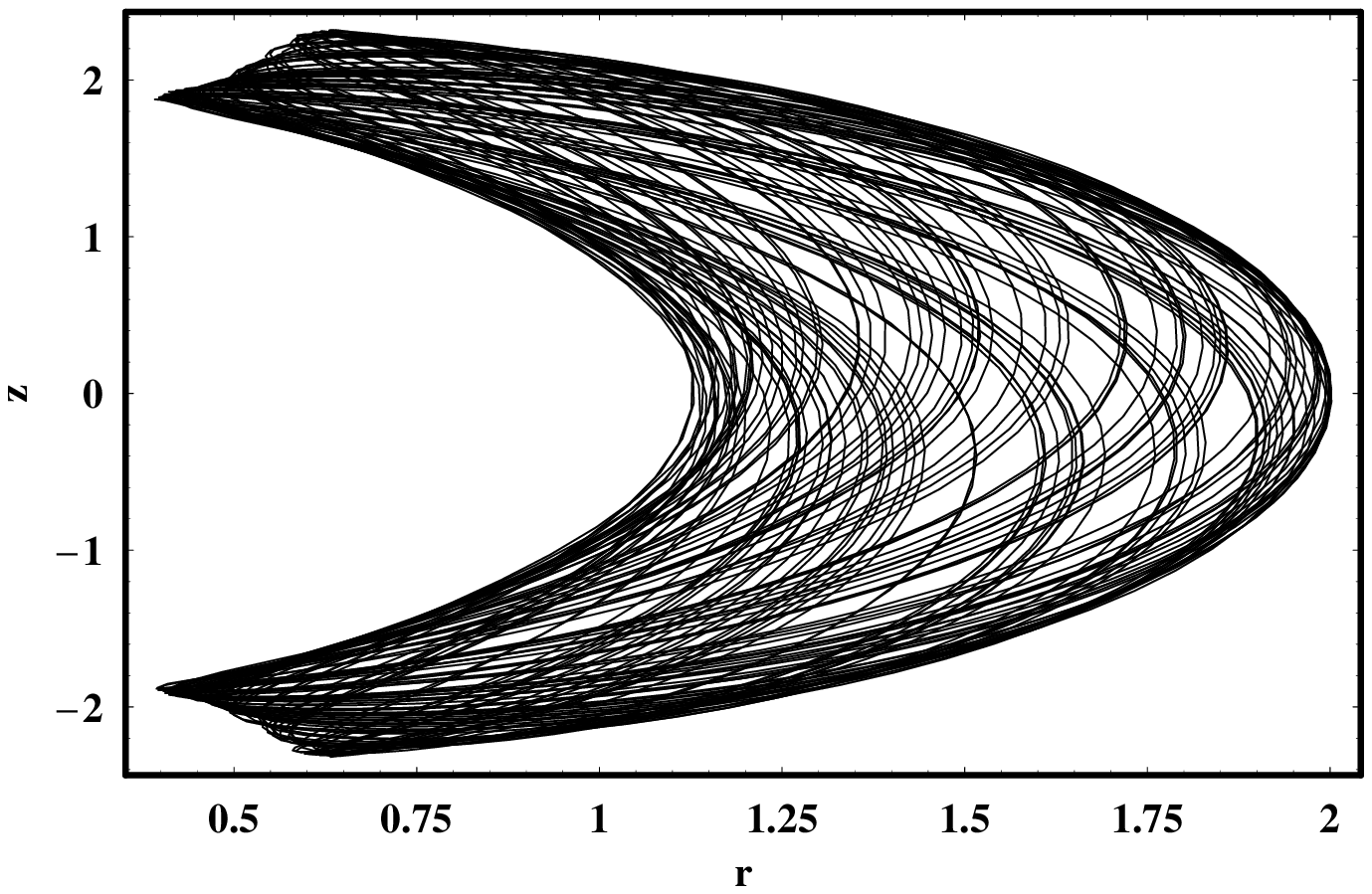}}\hspace{1cm}
                          \rotatebox{0}{\includegraphics*{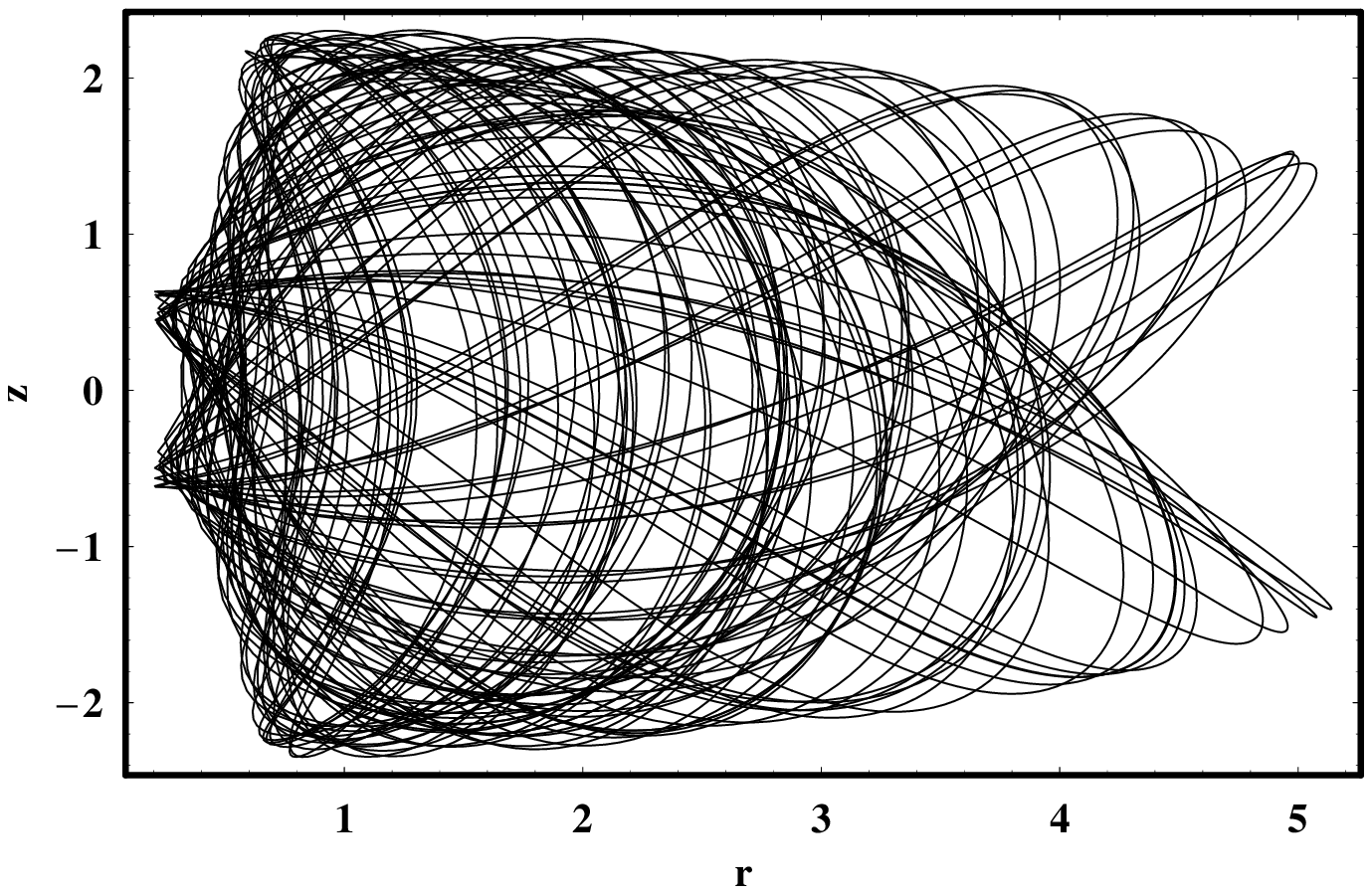}}}
\resizebox{0.95\hsize}{!}{\rotatebox{0}{\includegraphics*{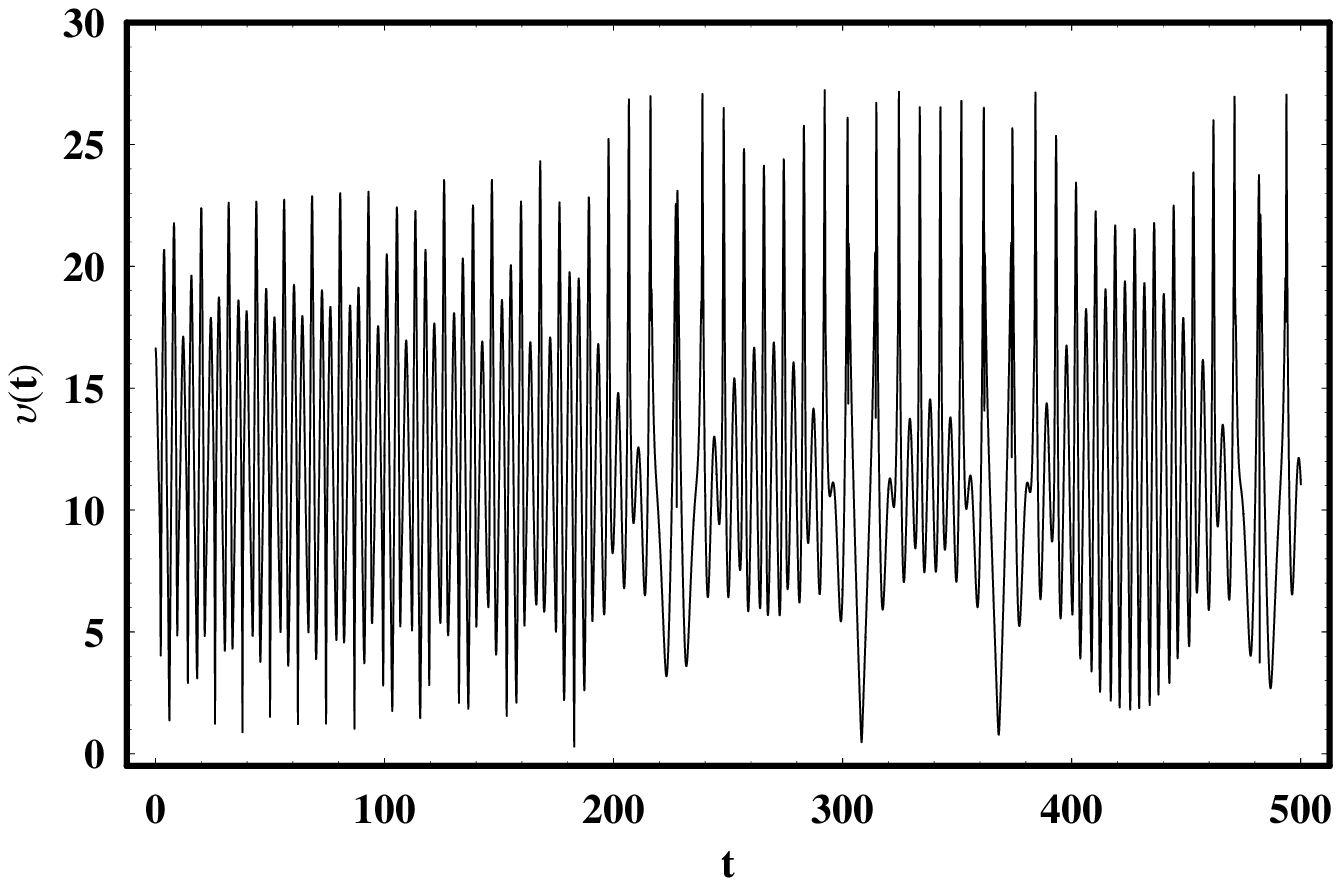}}\hspace{1cm}
                          \rotatebox{0}{\includegraphics*{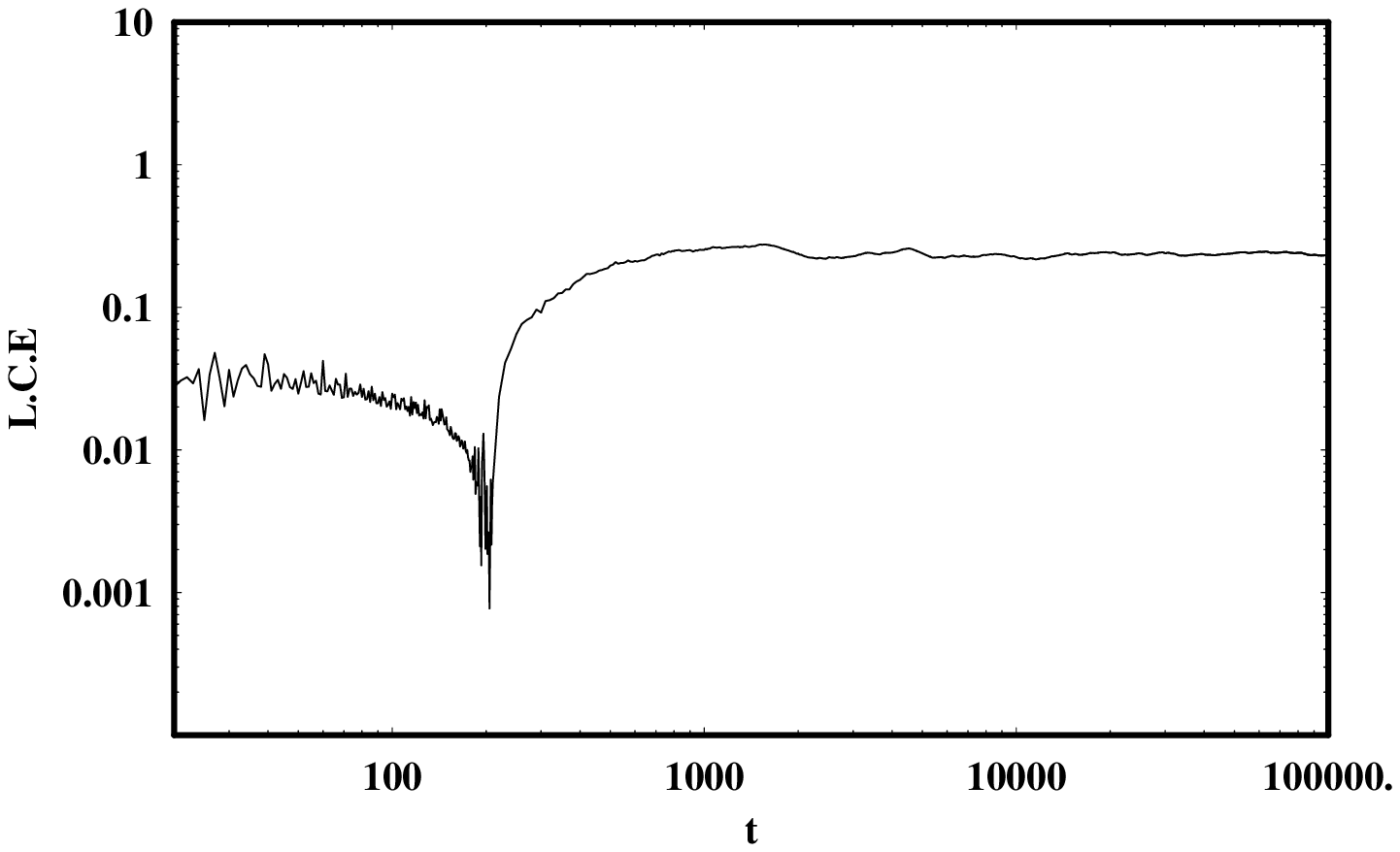}}}
\vskip 0.1cm
\caption{(a-d): (a-b): Evolution of an orbit as the external perturbation $\lambda$ changes with time, following the second of equations (21). The orbit starts as regular orbit and gradually becomes chaotic. (c-down left): Evolution of the total velocity of the orbit and (d-down right): Evolution of the L.C.E of this orbit. Details are given in the text.}
\end{figure*}

Figure 7a shows the evolution of an orbit, as the external perturbation $\lambda$ changes with time following the second of equations (21). The initial conditions are: $r_0 = 2.0, z_0 = p_{r0} = 0, \alpha = 1.9, c = 0.25$, while the initial value of energy is $E=300$. The value of angular momentum is $L_z = 5$. The initial value of $\lambda$ is 0, while $k_2$ is equal to 0.1. Figure 7a shows the orbit for the first 210 time units, while Figure 7b shows the orbit for the rest 90 time units. One observes, that the orbit starts as a regular orbit and gradually becomes chaotic, as the value of the external perturbation $\lambda$ increases. After 210 time units, the value of $\lambda$ becomes 21. At this point, the evolution of our dynamical system stops and the orbit runs for 90 more time units, with its new value of energy, which is now $E=275.38$. In Figure 7c we present the evolution of the total velocity $\upsilon (t)=\sqrt {p_r^2+p_z^2}$ of the test particle, as a function of time for the above orbit and for time interval of 500 time units. One observes that, at about $t = 200$ time units, the velocity profile changes, while at the same time the velocity increases. This indicates that, in practice, the orbit becomes chaotic after the external perturbation has reached the value $\lambda =20$. Here one must notice, that the above results are in agreement with observational data, where an increase of the stellar velocity is expected, in regions with significant chaos. Moreover, observations show, that in chaotic regions one expects to get an asymmetric velocity profile [13]. As one can see in Figure 7c, our velocity profile becomes asymmetric, when the motion changes from regular to chaotic. In order to double check our results, we computed the maximal Lyapunov Characteristic Exponent (L.C.E), for a time period of $10^5$ time units, which is shown in Figure 7d. The L.C.E indicates regular motion for the first 210 time units and then has a mean value, of about 0.25, which shows that the motion has become chaotic.

The above analysis shows, that the role of the external perturbation is not only to affect the nature of orbits (regular or chaotic), but also to change the profile of the velocities of the stars. All numerical calculations suggest, that as the external perturbation increases and the orbit tends to be chaotic, the velocity increases and its profile changes.

\section{Discussion and conclusions}

In this work we have used a simple potential, in order to study the dynamical behavior of a galaxy with an additional external perturbation, caused by a companion galaxy. Our aim was to investigate the consequences of the external perturbation on the character of motion (regular or chaotic). Moreover, we have tried to connect the strength of the external perturbation, the flattening parameter and the radius of the nucleus, with the conserved component of the angular momentum $L_z$.

The numerical calculations suggest, that strong external perturbations cause large chaotic regions on the phase plane. Spherical galaxies display smaller chaotic regions than flat ones. On the other hand, numerical and theoretical outcomes indicate that linear relationships exist between the critical value of the angular momentum $L_{zc}$ and the dynamical parameters of the system, that is the strength of the external perturbation $\lambda$, the flattening parameter $\alpha$ and the radius of the nucleus $c$. Moreover the extent of the chaotic regions observed in the $r-p_r$ phase plane, increases linearly, as the strength of the external perturbation and the flattening parameter increases. On the contrary, the percentage covered by chaotic orbits in the phase plane, decreases linearly as the scale length of the nucleus increases and becoming less dense.

The magnitude of the core radius $c$ together with the value of the angular momentum $L_z$, are two basic parameters for the dynamical system to display regular or chaotic motion. It was found numerically, that a linear relationship exists between the critical value of the angular momentum and the corresponding radius of the central concentration. The above mentioned relationship can also be obtained, using semi-theoretical methods.

An important role on the evolution of the chaotic motion is played by the flattening parameter $\alpha$ of the dynamical system. For a given value of the radius $c$ and the external perturbation $\lambda$, there is a linear relationship between $\alpha$ and the critical value of the angular momentum $L_{zc}$. This strongly suggests, that low angular momentum stars, display chaotic motion in highly flattened or perturbed elliptical galaxies, having a dense nucleus. Here we must note, that this behavior is similar to that observed in disk galaxies, studied by Caranicolas \& Innanen [3].

It is well known, from earlier work, that galaxies with dense and massive nuclei, produce chaotic regions [4]. The phenomenon
is strongly connected with the critical value of angular momentum $L_{zc}$ and the mass of the nucleus $M_n$. The corresponding relationship $[L_{zc}, M_n]$ was found to be linear [3]. Thus, we observe that the behavior of low angular momentum orbits in galaxies with massive and dense nuclei, is similar to the same orbits in galaxies with strong external perturbation. In both cases we observe large chaotic regions on the phase plane and the corresponding relationships $[L_{zc}, M_n]$, $[L_{zc}, \lambda]$, $[L_{zc}, \alpha]$ and $[L_{zc}, c]$ are linear.

It is important to note that, in order to observe chaos, the external perturbation must be combined with a dense nucleus. Numerical calculations not given here suggest that, when: $\lambda = 15, \alpha = 1.9, L_z = 10$, no chaotic motion is observed when $0.5 \leq c \leq 1$. Furthermore, it is well known that when $\lambda = 0$ [15], the system displays small chaotic regions, only for small values of $c$. The importance of this fact, can be shown as follows.

The density corresponding to potential (1) is
\begin{equation}
\rho (r,z) =  \frac{\upsilon_0^2}{4 \pi} \frac{\left[(\alpha + 2)c^2 +
\alpha \left(r^2 - (\alpha-2) z^2\right) \right]}{\left(r^2 + \alpha z^2 + c^2\right)^2}.
\end{equation}
For small values of $r$ and $z$, $\rho (r,z)$ tends to the value
\begin{equation}
\rho_0 = \frac{\upsilon_0^2\left(\alpha + 2\right)}{4 \pi c^2}.
\end{equation}
Thus, we see that the density near the center increases as $1 / c^2$. Equation (23) shows the critical role of the radius of nucleus in the character of motion. Therefore, we can say that our numerical experiments suggest that large external perturbations can cause significant chaos, if combined with high density objects in the central regions of galaxies.

An estimation of the total mass of the primary galaxy, can be obtained, if we assume a spherical galaxy of radius $R$. In this case $\alpha = 1$ and the density (22) becomes
\begin{equation}
\rho (w) = \frac{\upsilon_0^2}{4 \pi} \frac{\left( 3c^2 + w^2 \right)}{\left(w^2 + c^2\right)^2},
\end{equation}
where $w^2 = r^2 + z^2$. The mass of the galaxy is
\begin{equation}
M_G = 4 \pi \displaystyle\int^R_0 \rho(w) w^2 \,dw =
\upsilon_0^2 \left[ R - \frac{R c^2}{c^2 + R^2} \right].
\end{equation}
Taking $R = 20 kpc, c = 0.25 kpc$ we obtain the value: $M_G \simeq 4500$ mass units, that is $1.05 \times 10^{11} M_{\odot}$. Furthermore, we assume that the two bodies (the primary and the companion galaxy), are moving in circular orbits around the center of mass of the system, with an average period $T=5 \times 10^{9} yr$. The distance between the primary galaxy and its companion is $A=70 kpc$. The mass of the companion galaxy, can be obtained from the Kepler's third law
\begin{equation}
A \nu^2=G \left(M_G + M_c \right),
\end{equation}
where $\nu$ is the relative velocity of the two galaxies. From relation (26) and for the given values of the involving parameters, we have that $M_c \simeq 670$ mass units, which is equal to $1.55 \times 10^{10} M_{\odot}$. Our numerical results, can now be compared with observational data from the binary stellar system, consists of the giant elliptical galaxy $NGC$ $3379(M105)$ and its companion $NGC$ 3384.

Interesting results are found, in the case where the potential is time dependent. In this case, the numerical calculations indicate that the motion evolves from regular to chaotic, or from chaotic to regular depending on the particular values of the parameters $\alpha$ and $\lambda$. The main conclusion is that, as the external perturbation increases, the orbits become chaotic and the total velocity increases. Always we must remember, that all numerical calculation suggest that a large external perturbation, as described by our dynamical model (2), is responsible for producing a large amount of chaos, only if the mass density near the central regions is high.

Forty years ago, galactic activity and interactions between galaxies were viewed as unusual and rare. Nowadays, they seem to be segments in the life of many galaxies. From the astrophysical point of view, in the present work, we have tried to connect galactic activity and galactic interactions with the nature of orbits (regular or chaotic) and also with the behavior of the velocities of stars in the primary galaxy. We consider the outcomes of the present research , to be an initial effort, in order to explore matters in more detail. As results are positive, further investigation will be initiated to study all the available phase space, including orbital eccentricity of the companion and its inclinations to the primary galaxy.

\section*{Acknowledgments}

\textit{I would like to thank Professor N. D. Caranicolas for his fruitful discussions, during this research. I also would like to thank the anonymous referees for their very useful suggestions and comments, which improved the quality of the present paper.}

\end{document}